\documentclass{aa}

\usepackage{graphicx}
\graphicspath{{Figures/}}     
\usepackage{booktabs}
\usepackage{soul}
\usepackage{hyperref}
\usepackage{natbib}
\usepackage{float}
\usepackage{xcolor}
\usepackage{txfonts}

\begin{document}

   \title{Evolution of coronal mass ejections with and without sheaths from the inner to the outer heliosphere - statistical investigation for 1975--2022}

   \author{C. Larrodera
          \inst{1,2}\fnmsep\thanks{Corresponding author}
          \and
          M. Temmer \inst{2}
          }

   \institute{University of Alcalá,
              \\
              \email{carlos.larrodera@uah.es}
         \and
             University of Graz\\
             \email{manuela.temmer@uni-graz.at}
             }

   \date{Received ...; accepted ...}

  \abstract
  {}
   {This study covers a thorough statistical investigation of the evolution of interplanetary coronal mass ejections (ICMEs) with and without sheaths, through a broad heliocentric distance and temporal range. The analysis treats the sheath and magnetic obstacle (MO) separately to gain more insight about their physical properties. In detail, we aim to unravel different characteristics of these structures occurring over the inner and outer heliosphere.}
   {The method is based on a large statistical sample of ICMEs probed over different distances in the heliosphere. For this, information about detection times for sheath and MO from 13 individual ICME catalogs were collected  and cross-checked. The time information was then combined into a main catalog used as basis for the statistical investigation. The data analysis based on that covers a wealth of spacecraft missions enabling in-situ solar wind measurements from 1975--2022. This allows to study differences between solar cycles. }
{All the structures under study (sheath, MO with and without sheath) show the biggest increase in size together with the largest decrease in density at a distance $\sim$0.75 AU. At 1 AU we find different sizes for MOs with and without sheath, with the former being larger.
Up to 1 AU, the upstream solar wind shows the strongest pile-up close to the interface with the sheath. For larger distances the pile-up region seems to shift and recedes from that interface further into the upstream solar wind. 
his might refer to a change in the sheath formation mechanism (driven versus non-driven) with heliocentric distance, suggesting the relevance of the CME propagation and expansion behavior in the outer heliosphere. Comparison to previous studies shows inconsistencies over the solar cycle, which makes more detailed studies necessary to fully understand the evolution of ICME structures.}
   {}


   \maketitle

\section{Introduction} \label{sec:intro}

Coronal mass ejections (CMEs) are huge structures of plasma and magnetic field that are impulsively expelled from the Sun. The low plasma-beta structure, presumably a flux rope, drives the formation of other structures during its evolution through the ambient corona and interplanetary space. Close to the Sun, depending on the CMEs' initial speed, size and ambient coronal magnetic field, the so-called three-part CME is observed typically in white-light coronagraph data. It features a front region, a void and a center part \citep{Riley_2008, Mishra_2023}. More recent studies, applying 3D simulations and multi-spacecraft data, hint towards a two-front morphology consisting of a shock and piled-up sheath region \citep{Vourlidas_2013}. Indeed, from in-situ data more structures may be identified, including, for fast events, shock-sheath, leading edge, front, flux rope, and rear region \citep{Kilpua_2017,Temmer_2022}. The review by \citet{Wimmer_2006} further highlighted the existence of other specific regions, separated by discontinuities, and the center of the flux rope being a magnetic ejecta core.

The initial signature for identifying CMEs from in-situ data (referred to as Interplanetary CME; ICME) was established by \citet{Burlaga_1981} and \citet{Klein_1982}, focusing on magnetic field enhancements and the smooth rotation of the magnetic field. Subsequent studies applied alternative signatures for ICME detection. \citet{Richardson_1995} introduced the idea of combining the proton temperature with solar wind speed to calculate expected temperature values, specifically targeting the low temperature intervals characteristic of ICMEs. Expanding the range of signatures. \citet{Jian_2006} suggested the incorporation of total perpendicular pressure as a complementary variable to identify the presence of ICMEs. Additionally, the composition of ICMEs, which remains relatively constant after their departure from the Sun, has been employed as a proxy for their detection. \citet{Henke_2001} suggested as signature, the threshold of oxygen charge state ratio O$^{7+}$/O$^{6+}>$1, while \citet{Lepri_2001} and \citet{Lepri_2004} proposed average iron charge state values <Q>$_{Fe}>12$ as an identifying signature. An alternative approach for detecting ICMEs at Earth, proposed by \citet{Cane_2000}, relies on observing Forbush decreases, marked reductions in galactic cosmic ray intensities.
Due to the complexity of the ICME structures, the most practical approach is to use more than one signature \citep{Zurbuchen_2006} in order to reliably identify ICMEs in the solar wind \citep{Gosling_1997, Kilpua_2013}. In this research, most of the ICME identification processes employed in the catalogs, primarily concentrate on the magnetic field strength and the rotation of the magnetic field vector \citep{Mostl_2017,Mostl_2020,Catalog_2,Jian_2006} along with low proton temperature \citep{Catalog_2,Catalog_10b,Catalog_13} or the total pressure \citep{Jian_2006,Jian_2018}.

Under ideal circumstances, the expansion of the ICMEs in their propagation through the interplanetary medium would be compatible with self-similar expansion \citep{Farrugia_1993,Shimazu_2002,Demoulin_2008,Demoulin_2009a}, i.e. the velocity profile will show a linear decrease between the front and rear part of the magnetic obstacle. Nevertheless, the evolution of ICMEs with heliocentric distance is far from ideal. The inspection of different ICME structures in detail suggests that they have complex interactions with the ambient solar wind and evolutionary processes as they propagate through the interplanetary space. The interaction processes with the solar wind reveal several effects. One is the magneto hydrodynamic (MHD) drag, \citep{Cargill_2004,Vrsnak_2008,Vrsnak_2013} producing a kinematic adjustment to the ambient solar wind flow. Another effect is the pancaking of the frontal part of the CME, which leads to a deformation of the cross-section of the magnetic obstacle in the radial direction, producing a convex outward shape \citep{Hidalgo_2003,Ruffenach_2015}. Conditions favoring magnetic reconnection between the interplanetary magnetic field and the magnetic obstacle, produce a decrease in the ICME magnetic flux along with a decrease in the magnetic obstacle cross-section area \citep{Dasso_2007,Ruffenach_2012}. 

The ICME sheath region and its relation to the magnetic driver has gained interest in recent years \citep{DeForest_2013,Mitsakou_2014, Masias-Meza_2016, Kilpua_2017, Janvier_2019, Lugaz_2020, Salman_2020a, Salman_2021, Temmer_2022}. The sheath is typically defined as a shock-compressed, heated and turbulent material with a much larger plasma-beta than the magnetic obstacle. \citet{Das_2011} derived through the study of synthetic CMEs in the lower corona, that the magnetic field draping around the CME front creates a pile-up density compression region. \citet{DeForest_2013} and \citet{Lugaz_2020}, using in-situ measurements of ICMEs, detected the pile-up region and found that the sheath can be composed not only of compressed solar wind but also of coronal material. \citet{Siscoe_2008}, along with \citet{Salman_2021}, proposed a two-way formation mechanism for the sheath caused by the propagation and expansion of the magnetic obstacle through the solar wind. The characteristic feature of the propagation-dominated sheath is that the solar wind gets largely deflected sideways enabling to flow around the obstacle. On the other hand, the expansion-dominated sheath would refer to a continuous pile-up of solar wind all around the expanding magnetic obstacle \citep[see also][]{Siscoe_2008}. However, these would be very ideal situations, and pure propagation-dominated or pure expansion-dominated sheaths would be rare in terms of occurrence. Most likely, the formation mechanism of the sheaths might involve the combination of both mechanisms. Sheaths are therefore clearly related to the evolution of the magnetic obstacle that drives them, but also to the background solar wind in which they form. \citet{Temmer_2021} shows that the amount of piled-up material of the sheath, would depend on the density and solar wind flow speed and also on the magnetic obstacle size, since wider magnetic obstacles lead to stronger mass pile-up. The speed of the magnetic obstacle is also relevant for the sheath formation, as derived by \citet{Masias-Meza_2016}, who show that more massive sheaths are related to slow magnetic obstacles at 1 AU. \citet{Janvier_2019} study ICMEs at different heliocentric distances and determine that the median magnetic field magnitude in the sheath correlates well with the magnetic obstacle speed at 1 AU. As the background solar wind affects the sheaths, sheaths may also affect the magnetic obstacle that initially drives them and later follows them, and vice versa. 

In the present scientific study, our comprehensive analysis serves to corroborate and extend prior findings concerning the evolutionary patterns exhibited by magnetic ejecta as a function of heliocentric distance. Our investigation is conducted utilizing a dataset encompassing $\sim$2000 ICMEs between heliocentric distances of 0.25 and 5 AU. We not only extend the heliocentric distance analysis but also the time range of previous investigations by covering two entire solar cycles and the rising part of the SC25. Notably, limited scientific inquiries have been dedicated to study the sheath properties as a function of heliocentric distance or solar cycle. In this study, we address both of these aspects by delivering an exhaustive investigation into the sheath evolution, specifically focusing on variations in size and density. An additional facet of our research encompasses an exploration of how the properties of the upstream solar wind influence the propagation of ICME structures within the interplanetary medium. The investigation characterizes this interaction by assessing the ratio between the upstream solar wind and the various structural components within ICMEs. 

The paper is organized as follows, Section \ref{sec:data} details how we define the structures present in the ICMEs along with some statistics about the percentage distribution of ICMEs according to the heliocentric distance and solar cycle. Section \ref{sec:results} provides the results obtained, where Subsection \ref{sec:size_evolution} shows the differences in size between the substructures, whereas Subsection \ref{sec:density evolution} shows the differences in density along with the relation with the upstream solar wind. The evolution of the structures with solar cycles has been studied in Subsection \ref{sec:solar_cycle}, and how the magnetic obstacle would affect the sheath and vice versa is analyzed in Subsection \ref{sec:mo_sheath_ratio}. Finally, Sections~\ref{sec:conclusions} and \ref{Conclusion} summarize the results obtained from this research and draw some conclusions. 

\section{Data} \label{sec:data}

\begin{figure}[H]
	\includegraphics[width=0.8\columnwidth]{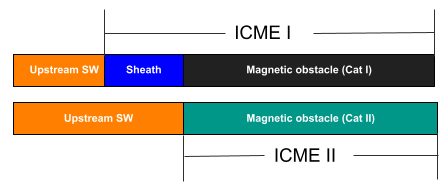} 
  \caption{Definition of ICME structures and ambient solar wind upstream of the ICME.}
  \label{fig:structures}
\end{figure}

The dataset used in this research is a combination of 13 in-situ measured ICME catalogs from different authors. Details on each catalog and references are given in the Appendix \ref{app:cat_references}. The dataset includes events which occurred from 1975 to 2022, i.e., fully covering SC21--24. After removing overlapping events, the dataset contains, 2136 separate ICMEs. The analysis is based on the in-ecliptic sample of data, with spacecraft locations between $-$10º and $+$10º. The 2003 in-ecliptic ICME events in the original catalogs are identified according to start and end times for sheath and magnetic obstacle (MO) regions. In this study, the dates are the only information inherited from the original catalogs, that allows us to classify the ICMEs into two types: ICMEs with a sheath region ahead of the MO (ICME I) and ICMEs without a clearly identified sheath region ahead of the MO (ICME II). Hence, ICME I consists of two structures, sheath and MO Cat I, while ICME II covers only one structure, the MO Cat II. The two ICME categories and upstream solar wind regions are visualized in Figure \ref{fig:structures}. 

During the time range between 1975 and 2022, several spacecraft missions were sampling ICMEs over different distances. Helios data cover 1976--1981 and distance range 0.3--1AU \citep{Scearce_1975}, Ulysses covers 1990--2007 and distance range 1.3--5.42AU \citep{Balogh_1992}. Located at 1AU: ACE 1997--active \citep{McComas_1998}, Wind 1994--active \citep{Ogilvie_1996}, STEREO 2006-active \citet{STEREO}, Parker Solar Probe 2018--active over 0.05--1AU \citep{Fox_2016} and Solar Orbiter 2020--active over 0.28--1AU \citep{Muller_2013}.Other spacecraft located in different planets have detected ICMEs, e.g. MAVEN \citep{MAVEN}, VEX \citep{VEX}, MESSENGER \citep{MESSENGER} or BepiColombo \citep{BColombo}. From these missions, we have obtained plasma and magnetic field measurements of the upstream solar wind, sheath and magnetic obstacle (see Figure \ref{fig:structures}). 

Table \ref{tab:icme_sc_distance} shows the information about the percentage distribution of ICMEs over the different solar cycles and heliocentric distance. We define 'Inner heliosphere' as distances $r<1$ AU and 'Outer heliosphere' as distances $r\ge1$ AU (cf.\,second column in Table \ref{tab:icme_sc_distance}).
Due to the lack of available spacecraft before SC23, SC20--22 cover only 5.7\% of ICMEs in the dataset. While SC23 and SC24 show a drop in ICME type II, the decrease in SC24 is slightly higher. Overall, the best data coverage is given for SC24 over the inner heliosphere. 

\begin{table}[H]
\begin{tabular}{ccccc}
\toprule
\textbf{Solar cycle}&\textbf{Region}&\textbf{ICME I} & \textbf{ICME II}& \textbf{TOTAL} \\ \toprule
SC20&Inner&0&0.3&0.3\\
SC20&Outer&0&0&0\\
SC21&Inner&0&4.4&4.4\\
SC21&Outer&0&0&0\\
SC22&Inner&0.4&0&0.4\\
SC22&Outer&0.1&0.5&0.6\\
SC23&Inner&13.4&6.7&20.1\\
SC23&Outer&7.4&5.5&12.9\\
SC24&Inner&22.6&11.8&34.4\\
SC24&Outer&7.6&5.2&12.8\\
SC25&Inner&7.1&5.3&12.4\\
SC25&Outer&1.0&0.7&1.7\\
\bottomrule
\end{tabular}
\caption{Percentage distribution (\%) of ICMEs distribution for each solar cycle and distance. Inner (r$<$1AU) and outer (r$\ge$1AU) heliosphere.}
\label{tab:icme_sc_distance}
\end{table}
\section{Results for in-ecliptic ICMEs} \label{sec:results}

\subsection{Size evolution with heliocentric distance} \label{sec:size_evolution}

For the 2003 in-ecliptic ICMEs, we have calculated their size (S) by multiplying the average speed by the duration, i.e., $S=\langle v\rangle \cdot \Delta t$. To study the size evolution of each structure as function of heliocentric distance, we present two approaches: (1) To be comparable with previous studies we fit the data with a power law function, $S=\delta\cdot r^{\beta}$, covering three distance ranges (the fitting parameters are detailed in Table \ref{tab:size_fitting_param}). (2) We calculate the statistical distribution of the size covering six distance ranges (detailed in Table \ref{tab:distance_range}).

Table \ref{tab:size_fitting_param} gives the power law fitting parameters for the three structures (sheath, MO Cat I and MO Cat II) in three different distance ranges: (1) inner \& outer (between 0.25 and 5.42 AU), (2) inner heliosphere ($r<1$ AU) and (3) outer heliosphere ($r\ge1$ AU). 72\% of the ICMEs are measured in the inner heliosphere, with 60\% being ICME I. On the other hand, 28\% ICMEs are measured in the outer heliosphere, of which 58\% are ICME I. In Table \ref{tab:size_fitting_param}, the structure labeled as $\overline{MO}$ refers to the average power law fitting parameters obtained from previous studies (a detailed review of these is given in Appendix \ref{app:size}). The different results show clear differences in the size and expansion behavior of the structures. 

\begin{table}[H]
\begin{tabular}{cccc}
\toprule
Distance (AU) &Structure& $\delta$ & $\beta$ \\  \toprule
0.25-5.42 & Sheath& $0.111\pm0.003$ & $0.795\pm0.369$\\
0.25-5.42 & MO Cat I& $0.269\pm0.005$ & $1.145\pm0.330$\\
0.25-5.42 & MO Cat II& $0.234\pm0.008$ & $0.600\pm0.037$\\  
Inner \& Outer & $\overline{MO}$ & $0.205\pm0.014$ & $0.795\pm0.086$\\ \midrule

0.25-0.99 & Sheath& $0.115\pm0.004$ & $1.719\pm0.736$\\
0.25-0.99 & MO Cat I& $0.275\pm0.008$ & $2.416\pm0.784$\\
0.25-0.99 & MO Cat II& $0.234\pm0.011$ & $0.805\pm0.236$\\  
Inner & $\overline{MO}$ & $0.250\pm0.050$ & $0.808\pm0.218$\\  \midrule

1-5.42 & Sheath& $0.111\pm0.005$ & $-0.657\pm1.106$\\
1-5.42 & MO Cat I& $0.283\pm0.010$ & $-0.35\pm0.853$\\
1-5.42 & MO Cat II& $0.242\pm0.021$ & $0.576\pm0.071$\\ 
Outer & $\overline{MO}$ & - & $0.630\pm0.427$\\ \bottomrule
\end{tabular}
\caption{Power law size fitting parameters ($S=\delta\cdot r^{\beta}$) for each structure and heliocentric distance. Each block represents a different distance range (Inner \& Outer, Inner, Outer) detailed in the first column. $\overline{MO}$ is the average value from previous studies, as detailed in Appendix \ref{app:size}.}
\label{tab:size_fitting_param}
\end{table}

The power law fitting parameters given in Table \ref{tab:size_fitting_param} suggest that the $\delta$ parameter, i.e. the size of the magnetic obstacle at 1 AU, is $\sim$15\% bigger for MO Cat I than MO Cat II. This result is independent of the heliocentric distance, revealing a small standard deviation of the data. On the other hand, the $\beta$ values of MO Cat I and II, i.e., the trend of the size with heliocentric distance are clearly dependent on the distance range chosen and reveal a high standard deviation. The evolution over heliocentric distance suggests that MO Cat I events undergo a stronger expansion, i.e., higher $\beta$, in comparison to MO Cat II events. 

The distance ranges chosen allow comparing our results to those from previous studies who used the same power law fitting method.  $\overline{MO}$ in Table \ref{tab:size_fitting_param} represents the average value and standard deviation of the power law fitting parameters as given in previous studies - we note that the sheath structures were not considered in most of these studies. According to the distance range sampled by the different authors, the definition of the distance ranges is Inner \& Outer (ca. 0.3--5.4 AU), Inner (ca. 0.1--1 AU) and Outer (ca. 1.4--5.4 AU). The comparison of our results shows that the MO size for both categories at 1 AU, is larger for the entire distance range, while a good match is obtained for the inner heliosphere (averaged for MO Cat I and II $\delta$=0.254). 

The derived sheath $\beta$ parameter reveals a large standard deviation in the outer heliosphere, however, for the dataset covering the entire as well as inner heliosphere the standard deviation is low. From this, we may conclude that the average value of $\beta$ indicates a sheath size evolution which is bigger than that derived from previous studies ($\beta=0.48$).

In order to analyze in more detail the evolution with heliocentric distance, we have binned the distance range into 6 bins of 0.25 AU width each. For better interpretation of the reliability of statistical results within each bin, Table \ref{tab:distance_range} shows the percentage of ICMEs in each distance interval. It should be noted that most of the ICMEs are located around 1 AU.

\begin{table}[H]
\begin{tabular}{ccccc}
\toprule
\textbf{Name} & \textbf{Distance (AU)}&\textbf{ICME I} & \textbf{ICME II}& \textbf{TOTAL} \\ \toprule
Dist 1& 0.25-0.49&3.8&2.2&6.0\\
Dist 2&0.50-0.74&3.8&4.4&8.2\\
Dist 3&0.75-0.99&35.8&22.0&57.8\\
Dist 4&1-1.24&16.0&8.9&24.9\\
Dist 5&1.25-1.49&0.2&0.3&0.5\\
Dist 6&$\ge$1.5&0&2.7&2.7\\
\bottomrule
\end{tabular}
\caption{Percentage distribution (\%) of ICMEs distribution for each distance interval.}
\label{tab:distance_range}
\end{table}

\begin{figure}[H]
	\includegraphics[width=\columnwidth]{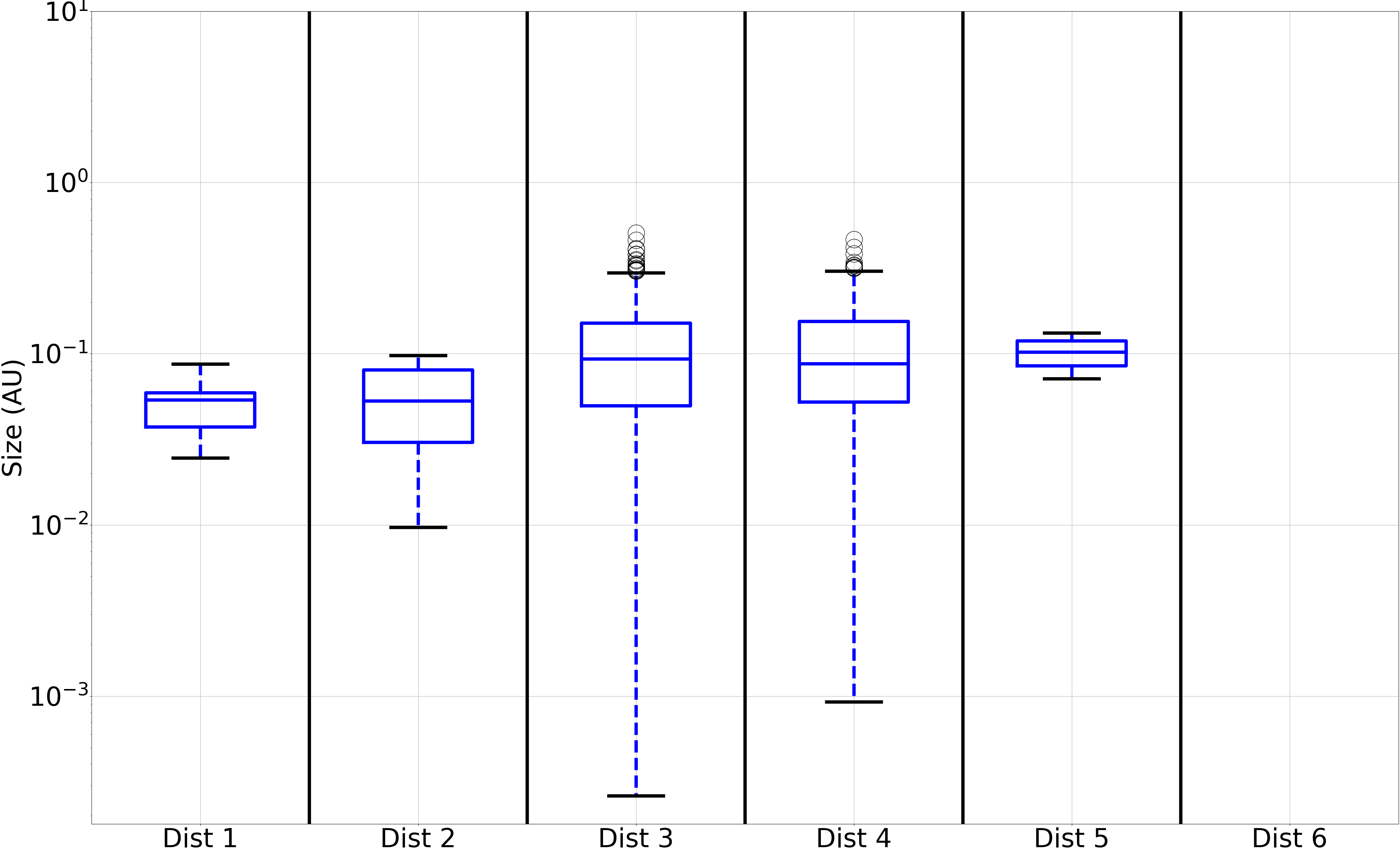} 
  \caption{Sheath size boxplots for the different distance intervals detailed in Table \ref{tab:distance_range}.}
  \label{fig:sh_size_evol}
\end{figure}

Figure \ref{fig:sh_size_evol} shows the boxplots for each distance interval detailed in Table \ref{tab:distance_range}. Each box provides the values of the first ($Q_{1}$), second ($Q_{2}$, median) and third quartile ($Q_{3}$) from the data distribution. The interquartile range (IQR) is the difference $Q_{3}-Q_{1}$. The upper and lower whiskers are defined as $Q_{3}+1.5\cdot$IQR and $Q_{1}-1.5\cdot$IQR, respectively. We observe a clear increase in sheath size from distance interval 2 ($r\ge0.5$ AU and $r<0.74$ AU) to distance interval 3 ($r\ge0.75$ AU and $r<0.99$ AU). Indeed, the size median value almost doubled from 0.05 AU to 0.09 AU. The IQR of distance interval 3 ($r\ge0.75$ AU and $r<0.99$ AU) and 4 ($r\ge1$ AU and $r<1.24$ AU) are comparable with the sheath size at 1 AU obtained from the power law fitting in the previous analysis. Beyond 1 AU (distance interval 4 onwards) the sheath size stays rather constant. We note that there are no ICME I reported for distance larger than 1.5 AU. This might be related to the distance where the ICME adjusts to the ambient solar wind flow, which reduces the number of detectable sheaths. On the other hand, we have a low coverage of ICME data in that region, since ICME I above 1.25 AU covers 0.2\% of the entire dataset.

\begin{figure}[H]
	\includegraphics[width=\columnwidth]{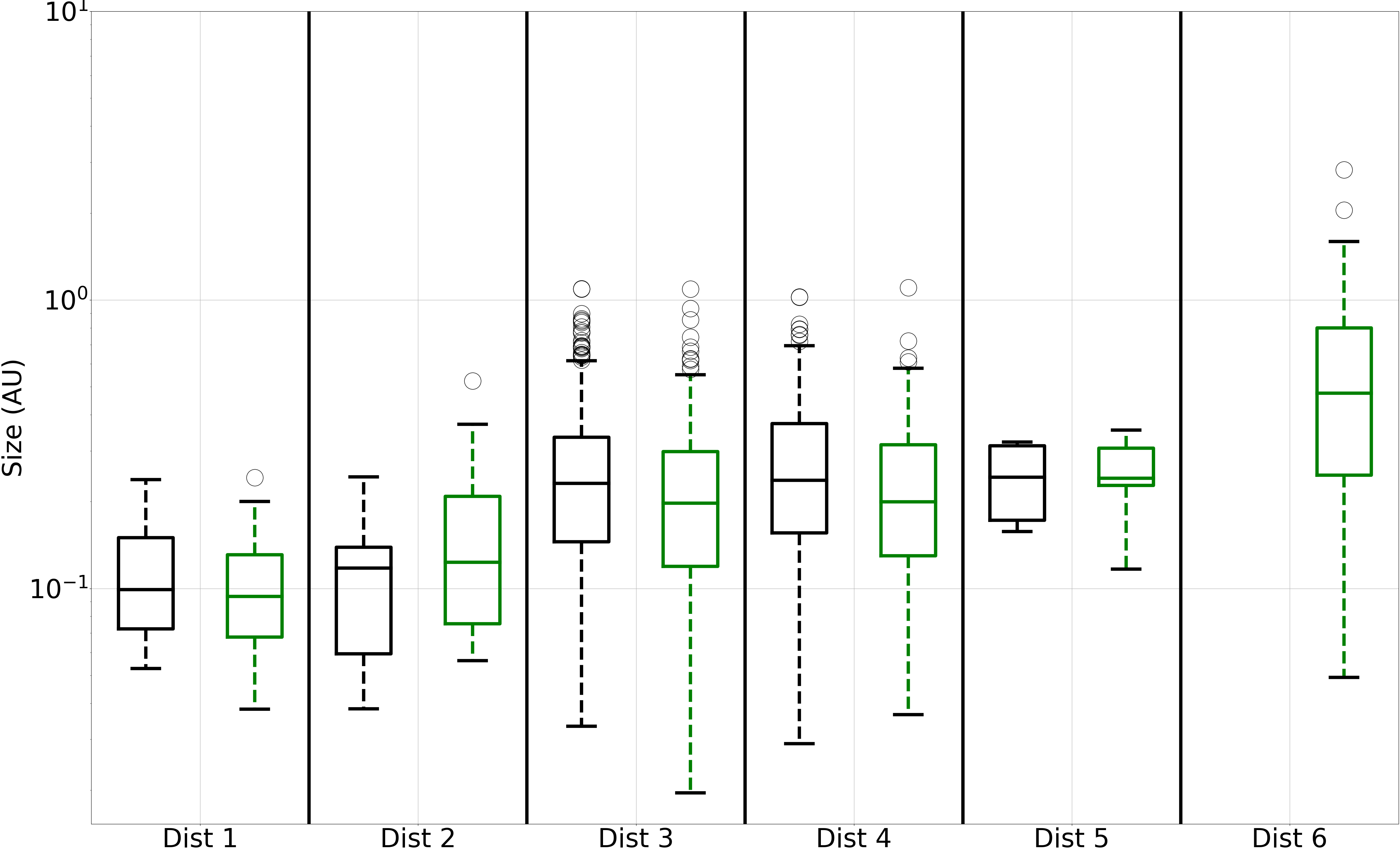}   
  \caption{MO Cat I (black) and MO Cat II (green) size boxplots for the different distance intervals detailed in Table \ref{tab:distance_range}.}
  \label{fig:mo_size_evol}
\end{figure}

Figure \ref{fig:mo_size_evol} shows size evolution of MO Cat I (black boxes) and MO Cat II (green boxes) in the six different distance intervals detailed in Table \ref{tab:distance_range}. The median sizes of both MO between 0.75 and 1.24 AU (distance interval 3 \& 4) are compatible with the results obtained from the power law fitting at 1 AU, reinforcing this result. The size of both MO categories show the same trend, a small increase in the most inner heliosphere (r$<$0.75 AU, distance intervals 1 \& 2) with an increase between 0.75 and 1 AU (distance interval 2 to 3), as occurs for the sheath. Beyond this distance, the size increases slowly, until r$>$1.5 AU (distance interval 6) where the size of MO Cat II  is almost double. We want to highlight that due to the small sample (2.7\% of the entire dataset) and the big IQR, this result should be taken with caution. The boxplot also shows that on average, the size of MO Cat II is lower than that of MO Cat I.

The strong expansion of MO Cat I might be an important factor in generating a sheath region. Results from observational data over various distance ranges such as \citet{Bothmer_1998, Leitner_2007, Demoulin_2008,Gulisano_2012,Vrsnak_2019} found that, on average, ICMEs tend to expand self-similarly, i.e., the size is directly proportional to the heliocentric distance (S$\propto$R). This self-similar expansion in the power law fitting parameters would be seen as $\beta\sim$1. Indeed, \citet{Gulisano_2012} found by analyzing magnetic clouds from Ulysses in the outer heliosphere, that non-perturbed ICMEs present an evolution compatible with a self-similar expansion. Nevertheless, when considering more complex situations such as magnetic obstacles perturbed by, e.g. a high speed stream or other magnetic obstacle, the evolution is no longer compatible with a self-similar expansion. Considering the separation between the inner and outer heliosphere, we derive that only MO Cat II expands self-similarly in the inner heliosphere ($\beta=0.805\pm0.236$) which is comparable to average results by other authors (cf.\, Table \ref{tab:size_fitting_param}). However, for the outer heliosphere, neither MO Cat I ($\beta=-0.35\pm0.853$) nor MO Cat II ($\beta=0.576\pm0.071$) expand self-similarly. The deviation from the self-similar expansion, as explained in \citet{Gulisano_2012,Vrsnak_2019}, could be related with the magnetic reconnection between the MO and the interplanetary magnetic field or the pancaking effect, i.e., the deformation in the shape of an initially circular MO, which is rather limiting the "apparent" radial expansion.
\subsection{Density evolution with heliocentric distance} \label{sec:density evolution}

The same methodology used for the size evolution has been applied to the density, i.e., (1) use a power law function to characterize the density evolution of the different structures over heliocentric distance, and (2) calculate the statistical distribution of the density covering the six distance intervals from Table \ref{tab:distance_range}.

Table \ref{tab:density_fitting_param} summarizes our results for the density evolution, providing the derived $\delta$ and $\beta$ parameters. To compare our results with those from previous works (see Appendix \ref{app:density} for details), we have averaged the fitting parameters from the power law function over the various distance ranges of Inner \& Outer (ca. 0.3--5.4 AU), Inner (ca. 0.1--1 AU) and Outer (ca. 1.4--5.4 AU). 

The $\delta$ of the MO Cat I and Cat II suggests that the density for both structures is very similar (between 7.1 and 7.3 cm$^{-3}$) and independent of the distance. The derived results are slightly higher than those reported in previous studies over the entire distance ranges available. For the density evolution over heliocentric distance ($\beta$) we find similar values for MO Cat I and II, in the range between $-$2.10 and $-$2.20 showing a decrease in density when moving from the inner to the outer heliosphere. 
In comparison, the average $\beta$ values from previous results ($\overline{MO}$) are lower, and derive an opposite trend for inner and outer heliosphere, i.e. an increase in density from the inner to the outer heliosphere.

The density of the sheath at 1 AU, i.e., $\delta$, is derived with an average value of $\sim$13 cm$^{-3}$ revealing to be independent of distance. Although the standard deviation of $\beta$ for the outer heliosphere is high ($\sim\pm$50\%), the other distance ranges have standard deviations for $\beta$ below 5\% ($\beta\sim$$-$2.1), suggesting that the decrease in density for the sheaths would be slightly lower than for the MO structures.

\begin{table}[H]
\begin{tabular}{cccc}
\toprule
Distance (AU) &Structure& $\delta$ & $\beta$ \\  \toprule
0.25-5.42 & Sheath& $13.03\pm0.34$ & $-2.14\pm0.06$\\
0.25-5.42 & MO Cat I& $7.26\pm0.18$ & $-2.10\pm0.06$\\
0.25-5.42 & MO Cat II& $7.22\pm0.33$ & $-2.07\pm0.05$\\  
Inner \& Outer & $\overline{MO}$  & $6.48\pm0.43$ & $-2.46\pm0.12$ \\
\midrule

0.25-0.99 & Sheath& $12.93\pm0.42$ & $-2.15\pm0.07$\\
0.25-0.99 & MO Cat I& $7.32\pm0.23$ & $-2.09\pm0.07$\\
0.25-0.99 & MO Cat II& $7.23\pm0.46$ & $-2.07\pm0.07$\\  
Inner & $\overline{MO}$  & $7.17\pm1.51$ & $-2.42\pm0.46$\\
\midrule

1-5.42 & Sheath& $13.56\pm0.65$ & $-3.67\pm1.93$\\
1-5.42 & MO Cat I& $7.11\pm0.29$ & $-2.22\pm1.38$\\
1-5.42 & MO Cat II& $7.21\pm0.33$ & $-2.22\pm0.78$\\ 
Outer & $\overline{MO}$  & - & $-1.78\pm0.53$\\
\bottomrule
\end{tabular}
\caption{Power law density fitting parameters ($N=\delta\cdot r^{\beta}$) for each structure and heliocentric distance. Each block represents a different distance range (Inner \& Outer, Inner, Outer) detailed in the first column. $\overline{MO}$ is the average value from previous studies, as detailed in Appendix \ref{app:size}.}
\label{tab:density_fitting_param}
\end{table}

\begin{figure}[H]
	\includegraphics[width=\columnwidth]{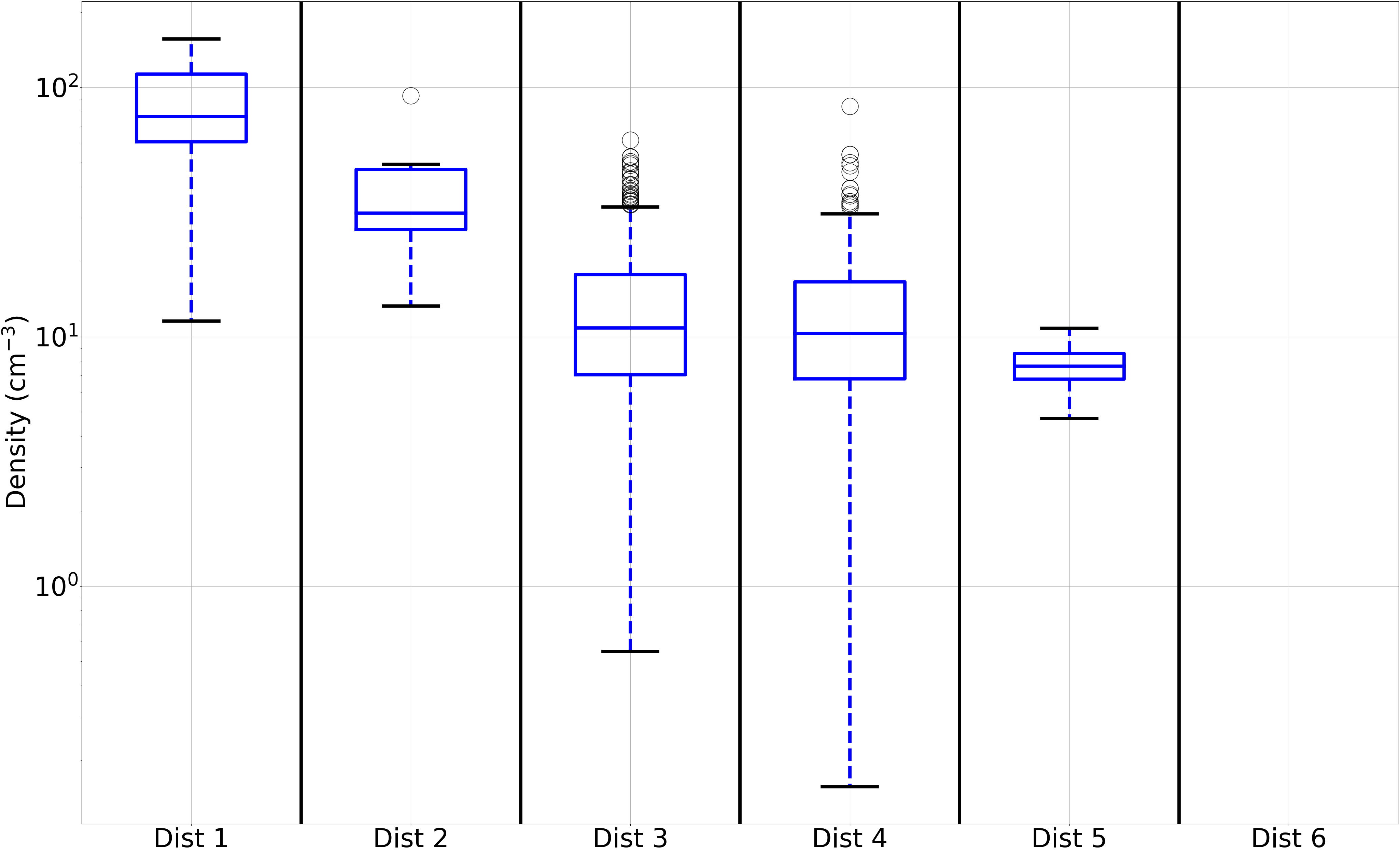} 
  \caption{Sheath density boxplots for different distance interval detailed in Table \ref{tab:distance_range}.}
  \label{fig:sh_density_evol}
\end{figure}

As we did for the size, in order to analyze in more detail the density evolution we split the heliocentric distance in six slices (see Table \ref{tab:distance_range}) to obtain the boxplots, which characterized the probability distribution function of the dataset as explained before. Figure \ref{fig:sh_density_evol} shows the density boxplots for the sheath structure, where we find that the median density observed in between 0.75 and 1.25 AU (distance interval 3 \& 4) are comparable with the density observed at 1 AU as obtained from the power law fitting. The sheath density evolution shows again a stronger decrease in the most inner heliosphere, below 0.75 AU (distance interval 1 \& 2). From this point onwards, the density continues to decrease but at a smaller rate, followed by another drop within distance interval 5. Note that no sheath structures were detected for heliocentric distance $\ge$1.5 AU (distance interval 6).

\begin{figure}[H]   
    \includegraphics[width=\columnwidth]{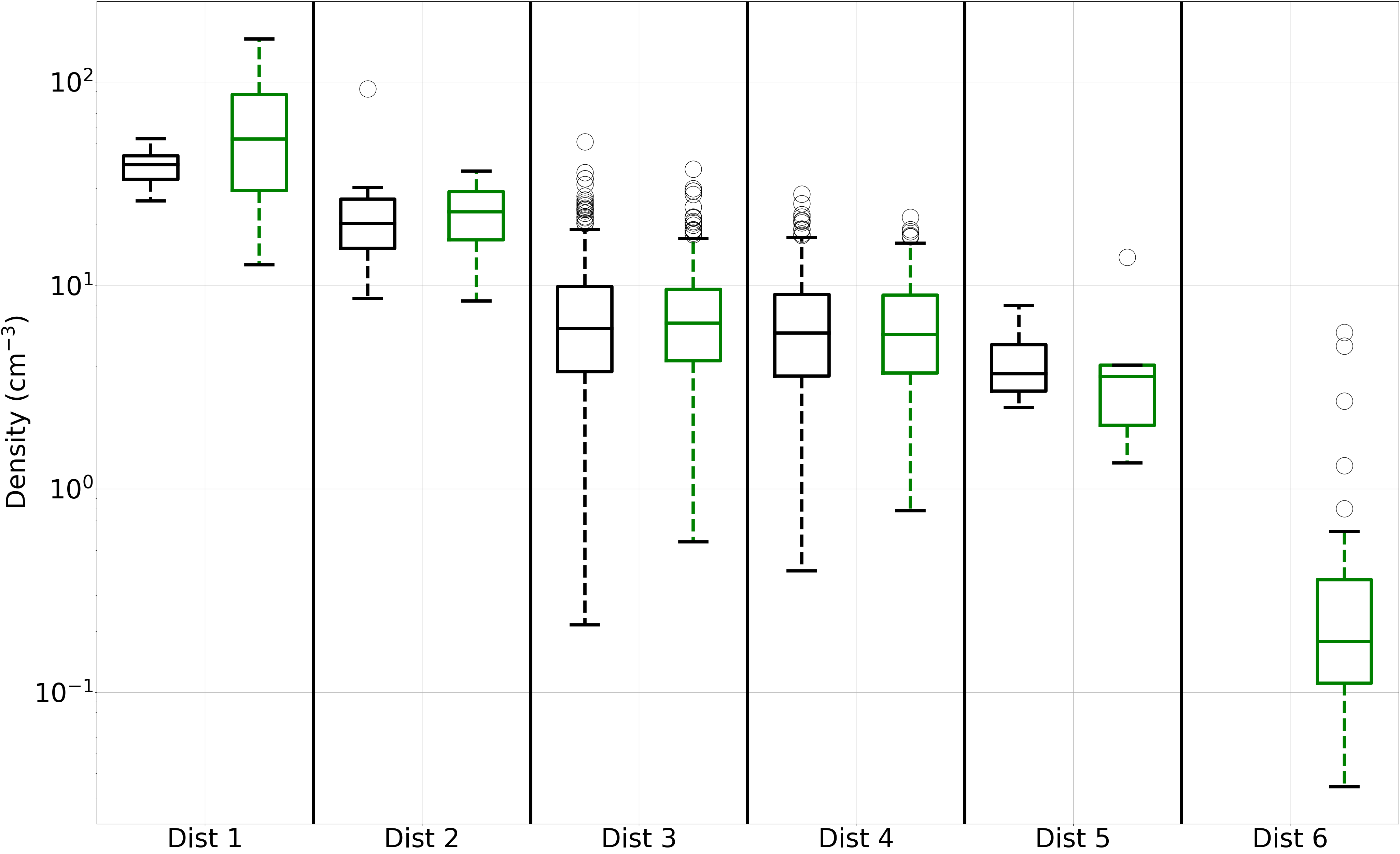}    \caption{MO Cat I (black) and MO Cat II (green) density boxplots for different distance interval, detailed in Table \ref{tab:distance_range}.}
  \label{fig:mo_density_evol}
\end{figure}

Figure \ref{fig:mo_density_evol} shows the results for the density evolution of the magnetic obstacles. The median MO density measured between 0.75 and 1.25 AU (distance interval 3 \& 4) is again compatible with the density derived at 1 AU as obtained from the power law fitting. Furthermore, we derive that the density median values for both MO Cat I and Cat II are very similar, except below 0.50 AU (distance interval 1) where MO Cat II shows higher density than MO Cat I. The trend in the density evolution from 0.5 AU onwards (distance interval 2) is for both MO structures similar, and comparable to the trend derived for the sheath density evolution. This is, a steeper density decrease in the most inner heliosphere, followed by a more gentle decrease beginning at 0.75 AU (distance interval 3) with another stronger decrease observed at 1.25 AU onwards (distance interval 5).

\subsubsection{Density ratio between upstream solar wind and structures} \label{sec:upstream}

\begin{figure}[H]
	\includegraphics[width=1.\columnwidth]{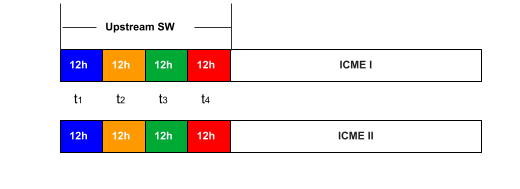}  
  \caption{Definition of upstream solar wind windows under investigation.}
  \label{fig:windows}
\end{figure}

The solar wind dynamical behavior affects the ICMEs in their propagation behavior. During the propagation through the interplanetary medium, the ICME is embedded in the ambient solar wind and interacts with it. Indeed, under which conditions strong sheath regions are generated and how they impact the evolution of the ICME is not yet fully understood. In the following, we use the extensive catalog presented in this study to make a statistical approach focusing on the density. To derive the average density of the upstream solar wind we extract a 48-hour window ahead of the ICME and separate it in 4 windows (t$_{1}$, t$_{2}$, t$_{3}$, and t$_{4}$) of 12 hours duration each (see Figure \ref{fig:windows}). 

We investigate how the upstream solar wind (SW) affects the sheath and the magnetic obstacle that interacts more directly with the solar wind, namely MO Cat II. For that we calculate the density ratios (RT) between the upstream solar wind, with the sheaths (SH) and the MO Cat II (MOII), given by $\rm RT_{SH}=\frac{N_{SH}}{N_{SW}}$, and $\rm RT_{MOII}=\frac{N_{MOII}}{N_{SW}}$, respectively. To cover the evolution of the ratio over heliocentric distance we use four distance intervals (1) $r\in \left[0.25,0.7\right)$ AU, (2) $r\in \left[0.7,1\right)$ AU, (3) $r \in \left[1,1.5\right)$ AU, and (4) $r\ge1.5$ AU. 

\begin{figure}[H]
	\includegraphics[width=\columnwidth]{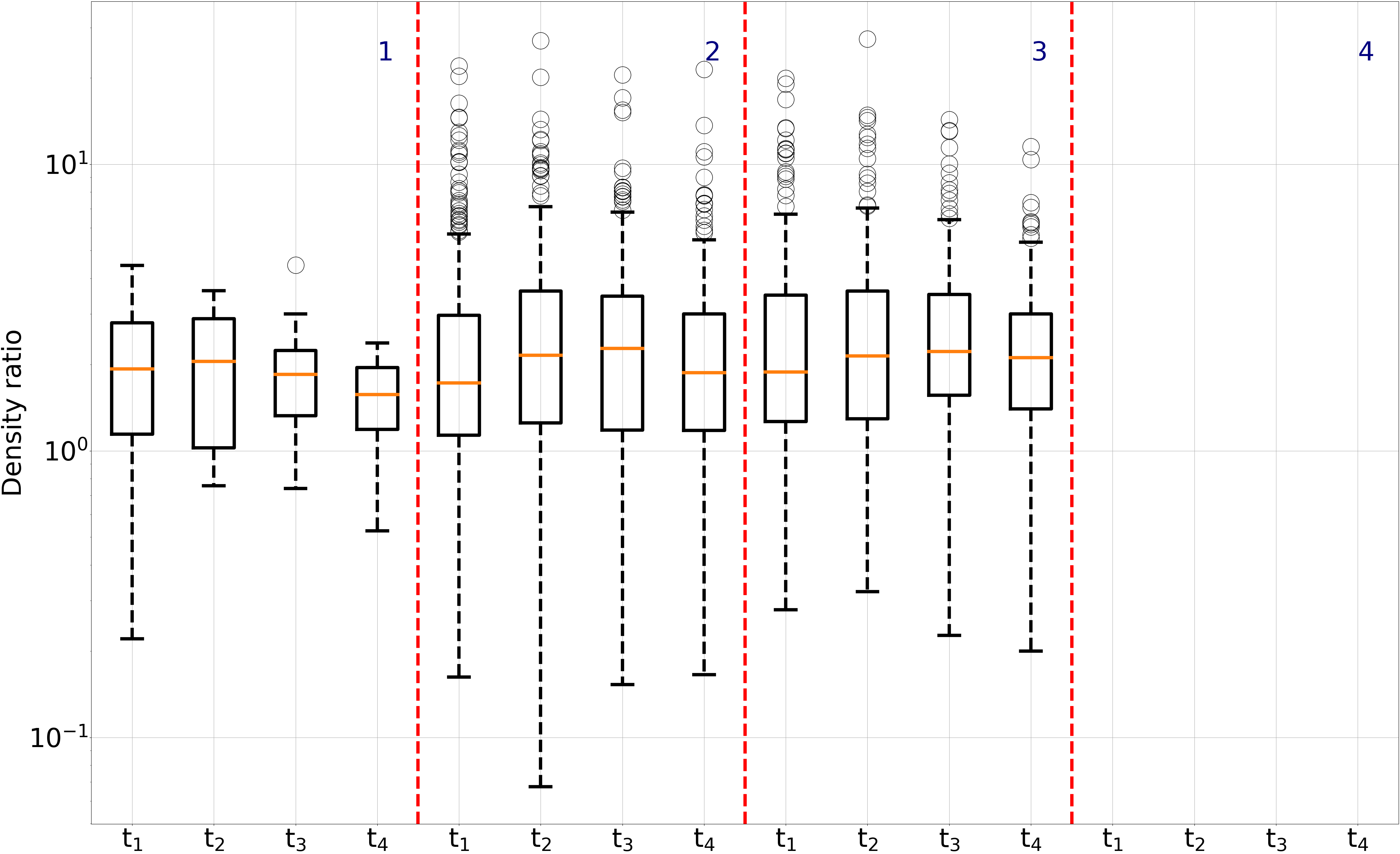}    
  \caption{Density ratio between sheath and upstream solar wind according to the time windows given in Figure \ref{fig:windows}. Each numbered block refers to a distance interval, as detailed in the text. Each box represents the result averaged over the 12 hours time window.}
  \label{fig:density_ratio_sheath}
\end{figure}

Figure \ref{fig:density_ratio_sheath} shows the density ratio of the sheath ($\rm RT_{SH}$). Each block (labeled from 1 to 4) represents a distance interval, as detailed before, and covers four boxplots of the sheath density ratio, with the same characteristics as given in Figure \ref{fig:sh_size_evol}. For all distance intervals, we obtain a median ratio bigger than 1, meaning that the density of the sheath is always larger than the density of the upstream solar wind. For distance interval 1 we observe a decreasing trend in the median density ratio from t$_{1}$to t$_{4}$. This behavior reflects a density increase from the farthest part of the upstream solar wind under investigation (t$_{1}$) towards the interface between solar wind and sheath (t$_{4}$). This trend changes when moving to the outer heliosphere, between 0.7 and 1.5 AU (distance intervals 2 and 3). At these distance ranges the region with highest density seems to shift into the upstream solar wind (t$_3$ and t$_2$) moving away from the interface between upstream solar wind and sheath (t$_4$).

\begin{figure}[H]
	\includegraphics[width=\columnwidth]{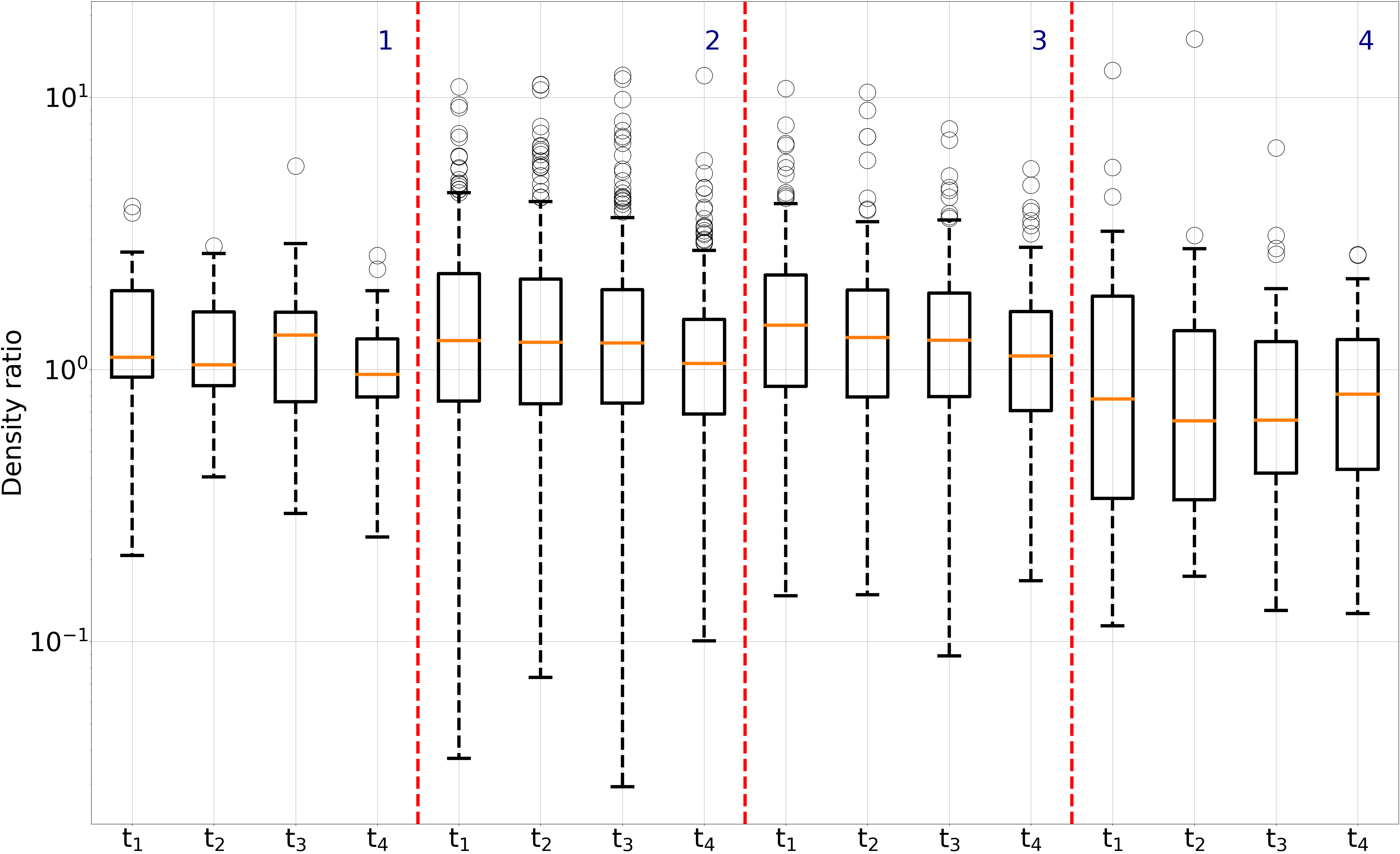}  
  \caption{Same as Figure~\ref{fig:density_ratio_sheath} but for the density ratio between MO Cat II and upstream solar wind.}
  \label{fig:density_ratio_mo_II}
\end{figure}

Figure \ref{fig:density_ratio_mo_II} provides the density ratio of the MO Cat II ($\rm RT_{MOII}$), which has a direct interaction with the ambient solar wind, as there is no clear sheath detected according to the catalogs. Overall, we find for distance intervals between 0.25 and 1.5 AU (block 1--3) density ratios greater than 1, whereas for heliocentric distances greater than 1.5 AU (block 4) the ratio density is lower than 1. Hence, the density of the MO Cat II in the outer heliosphere (r$\ge$1.5 AU) is lower than the upstream solar wind, reflecting the expansion behavior of the magnetic structure. In comparison to the sheath, we do not find a clear trend of the density ratio behavior from t$_{1}$ to t$_{4}$ between 0.25 and 0.7 AU (within distance interval 1). Between 0.7 and 1.5 AU (distance intervals 2 and 3) the density ratio shows a small decrease from t$_{1}$ to t$_{3}$ and a bigger one from t$_{3}$ to t$_{4}$, meaning an increase in density in the nearest region to the (presumable) interface between solar wind and MO Cat II. A possible explanation for this increase could be the presence of a sheath, undetected due to its weak characteristics. The trend above 1.5 AU (distance interval 4) suggests a higher upstream solar wind density in t$_{2}$ and t$_{3}$, i.e. the central regions under study, on the contrary to the trend for lower heliocentric distances. It should be noticed that, for the entire heliocentric distance, the IQR of the boxplots is compatible with 1, meaning, the density of the upstream solar wind and the structure would be similar.
\subsection{Structures and their evolution with solar cycle} \label{sec:solar_cycle}

The extensive catalog covers a large time range, allowing to investigate the ICME structure evolution over different solar cycles (SC20--SC25). Due to the number of spacecraft available, the ICME sampling is not homogenous over time, and the percentage of ICMEs for SC20, SC21 and SC22 is only 5.7\% (c.f. Table \ref{tab:icme_sc_distance}). Therefore, we focus our analysis on SC23, SC24 and the rising phase of SC25 representing 94.3\% of the ICMEs and covering more than 25 years from 1996 to 2023. With that we extend the results given by \citet{Kilpua_2017} about the SC24 and part of the SC25, and moreover, we introduce the differentiation of the results between the inner and the outer heliosphere.

\begin{figure}[H]
	\includegraphics[width=0.9\columnwidth]{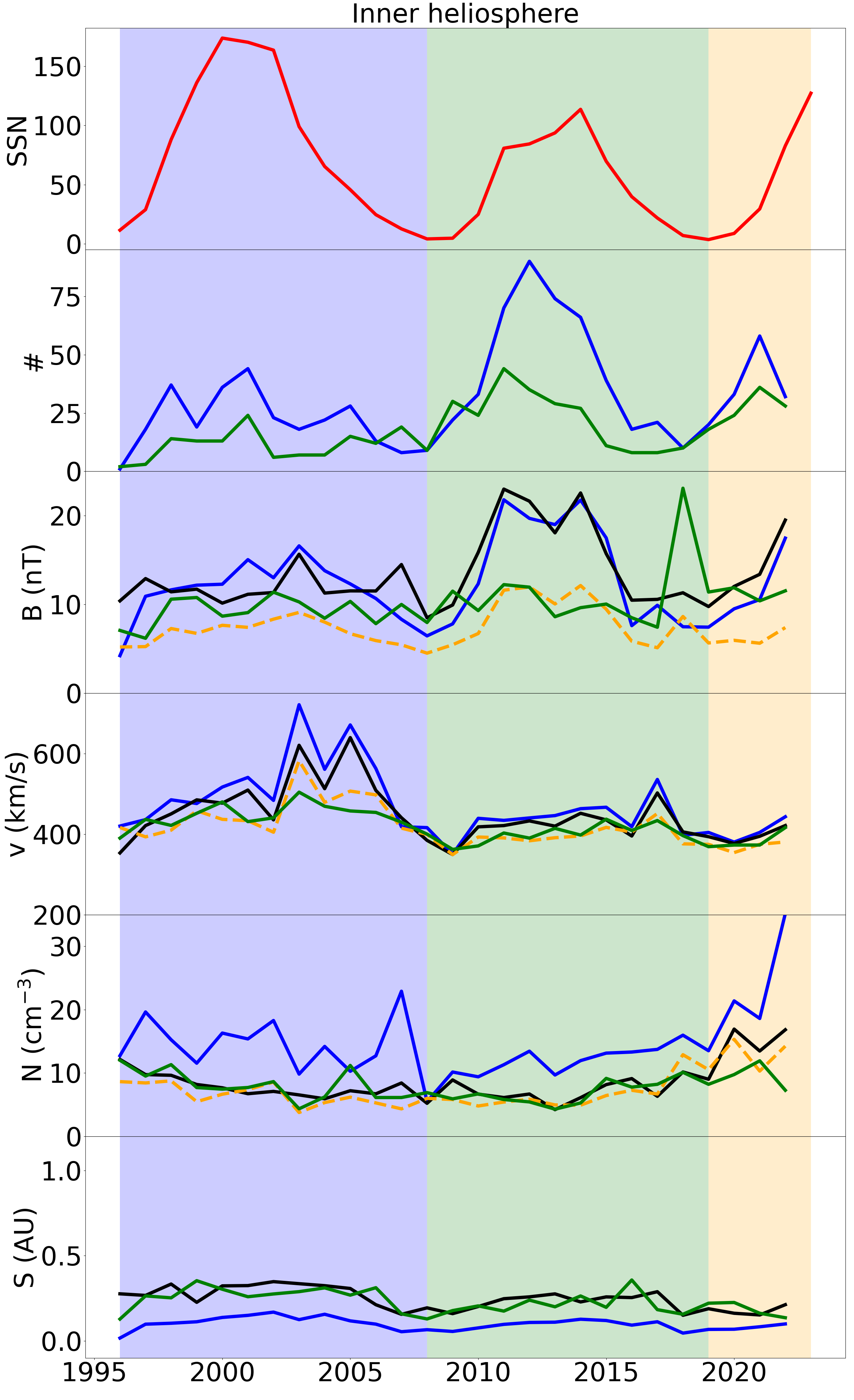}  
 \caption{Inner heliosphere yearly average values. The vertical color bands (blue, green and yellow) represent the SC23, 24 and part of SC25 respectively. From top to bottom, sunspot number (SSN), number of events (\#), magnetic field (B), speed (v), density (N) and size (S). The red line is the SSN, while the blue, black and green lines are the data for the sheaths, MO Cat I and MO Cat II respectively. The orange dashed line is the upstream solar wind yearly values. Since the number of sheaths and MO Cat I are the same, the second panel shows only the number of sheaths (blue line). }
 \label{fig:sc_inner}
\end{figure}

\begin{figure}[H]
	\includegraphics[width=0.9\columnwidth]{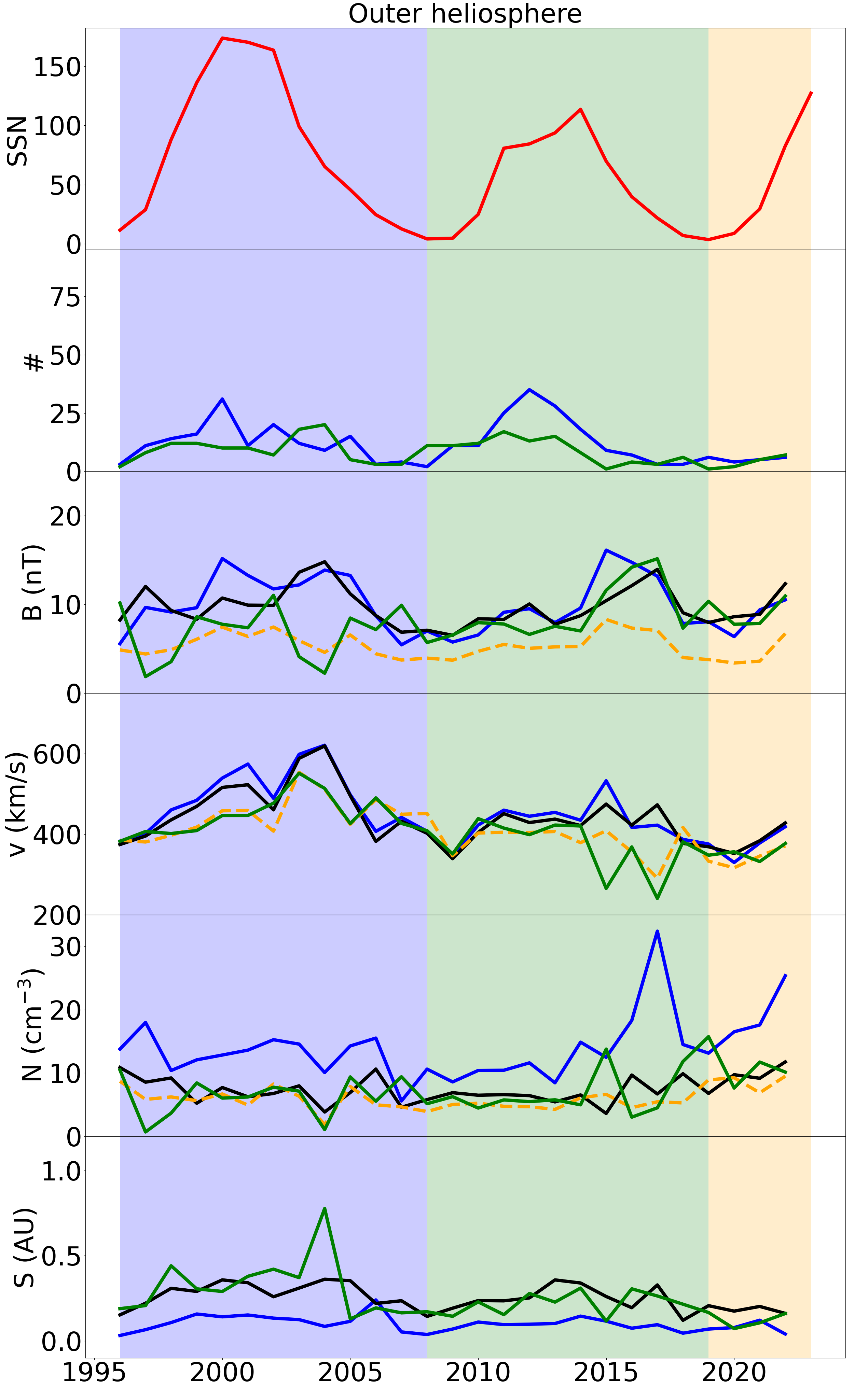} 
 \caption{Outer heliosphere yearly average values. The vertical color bands (blue, green and yellow) represent the SC23, 24 and part of SC25 respectively. From top to bottom, sunspot number (SSN), number of events (\#), magnetic field (B), speed (v), density (N) and size (S). The red line is the SSN, while the blue, black and green lines are the data for the sheaths, MO Cat I and MO Cat II respectively. The orange dashed line is the upstream solar wind yearly values. Since the number of sheaths and MO Cat I are the same, the second panel shows only the number of sheaths (blue line). }
 \label{fig:sc_outer}
\end{figure}

Figures \ref{fig:sc_inner} and \ref{fig:sc_outer} show the yearly averaged evolution of various ICME parameters for the inner and outer heliosphere, respectively, between 1996 and 2023. The colors of the lines are the same as in Figure \ref{fig:structures}, blue for the sheaths, black for MO Cat I and green for MO Cat II. For comparison, we have added the yearly averaged values of the upstream solar wind parameters, given as orange dashed lines. The color shaded regions represent the different solar cycles, blue SC23, green SC24 and yellow SC25.

For the inner heliosphere (Figure \ref{fig:sc_inner}) we see that the number of sheaths and MO Cat I\footnote{\label{note1}Note that the number of sheaths and MO Cat I are the same} (blue line) in comparison with MO Cat II (green line) is bigger for the solar cycles under study. During the maximum phase of SC23 the difference is 45\% and this difference increases to 51\% for SC24. In SC23, the three structures show a magnetic field higher than the average solar wind magnetic field (orange line). The sheaths (blue line) and MO Cat I (black line) show higher magnetic field than MO Cat II (green line) during the declining phase of SC23. In SC24, there is a clear difference between the magnetic field of sheath and MO Cat I in comparison with MO Cat II. In SC23 and SC24, the difference between MO Cat I and the upstream solar wind is 42\% and 48\% respectively. The yearly averaged speed of the structures are rather comparable over the solar cycles, although in the declining phase of SC23 there is a speed increase for the sheath and MO Cat I, while the MO Cat II events do not show such a clear trend and even fall below the upstream solar wind flow speed. In SC24 this increase also occurs, but it is smaller for the three structures, and again the MO Cat II speed is lower than the upstream solar wind speed. The density reflects the higher density of the sheaths in comparison with MO Cat I and MO Cat II in SC23 and SC24. Both MO categories have a density similar to that of the upstream solar wind in both solar cycles. On the other hand, a differentiation in density appears for both MO during the rising phase of SC25, but this result should be treated carefully due to the low sample for SC25. Lastly, the size of both MO remains almost constant in both solar cycles and is also bigger than the sheath size, which also remains constant over the solar cycles. 

For the outer heliosphere (Figure \ref{fig:sc_outer}) we obtain rather different results than for the inner heliosphere. In SC23, the number of sheaths and MO Cat I (blue line)\footref{note1} in comparison with MO Cat II (green line) in some phases are similar. Only in SC24, around the maximum, the number of sheaths are twice as numerous as MO Cat II. Regarding the magnetic field, there are no clear trends, although the magnetic field of sheaths (blue line) and MO Cat I (black line) seem to decrease during the declining phase and are bigger than the upstream solar wind magnetic field, while the MO Cat II during SC23 vary with values above and below the upstream solar wind magnetic field. During the declining phase of both the solar cycles and also during the rising phase of SC23, the sheath (blue line) and MO Cat I (black line) speed is bigger than the speed of MO Cat II (green line), while during the minima, the three structures present almost the same speed. By contrast to what occurs in the inner heliosphere the speed difference over the solar cycles are rather small. The density does not exhibit a clear dependency on the solar cycle. As expected, we observe higher densities in the sheaths compared to MO Cat I and Cat II, which have similar densities to the upstream solar wind. The size of the sheath (blue line) is smaller than that of both MO categories. The size of MO Cat II (green line) exhibits an increase during the ascending phase and reaches its maximum during SC23. Indeed, \citet{Gopalswamy_2014} through the correlation analysis between the speed and angular width of coronal mass ejections, found that during SC24, the CMEs are wider than those in SC23 due to the reduction in the total pressure by $\sim$ 40 \%. \citet{Gopalswamy_2015_2} compared the properties of the magnetic obstacles in SC23 and SC24, supporting the idea of a weaker SC24, showing also that the magnetic obstacles in SC23 are bigger than in SC24.

Several authors have studied the properties of ICMEs at 1 AU through different years \footnote{A detailed review of the results from previous authors can be found in the Appendix \ref{app:sc_evol}}, but these results show some discrepancies. During SC23, the results of the sheath density from \citet{Gopalswamy_2015_2} provide the highest values, followed by \citet{Mitsakou_2014} and \citet{Yermolaev_2021}. Results for the magnetic field of sheath structures during SC23, are found to be highest from \citet{Gopalswamy_2015_2}, which is in agreement with \citet{Kilpua_2017}, and are reported as almost twice as large as that from \citet{Mitsakou_2014} and \citet{Yermolaev_2021}. The results are also different for the sheath speed during SC23, for which \citet{Gopalswamy_2015_2} and \citet{Kilpua_2017} provide higher values than \citet{Mitsakou_2014} and \citet{Yermolaev_2021}. These discrepancies are also reflected in the results of the magnetic field strength for the MOs during SC23, for which \citet{Yermolaev_2021} give values almost three times lower than that from \citet{Gopalswamy_2015_2} and almost two times lower than \citet{Mitsakou_2014}. The results of SC24 also differ for \citet{Gopalswamy_2015_2} and \citet{Yermolaev_2021} for sheath and magnetic obstacle, although the density of the sheath is quite similar in this case. These results highlight the difficulties in finding an agreement on the mean values for each solar cycle. We do not provide the mean values for the near-Earth orbit, but for the inner and outer heliosphere. Tables \ref{tab:sheath_sc}, \ref{tab:moI_sc} and \ref{tab:moII_sc} show the average values calculated from our dataset for each solar cycle separately for sheaths, MO Cat I and MO Cat II, respectively. The first block in each table are the results considering the entire dataset, without filtering for inner, i.e., r$<$1 AU (second block) or outer, i.e. r$\ge$1 AU) (third block) heliosphere.

\begin{table}[H]
\begin{tabular}{ccccc}
\toprule
SC &N (cm$^{-3}$)& v (km/s) & B (nT) & Size (AU)\\  \toprule
23 & $14.0\pm9.7$ & $530.2\pm143.9$& $12.5\pm6.2$ & $0.128\pm0.089$\\  
24 & $11.7\pm8.6$ & $442.6\pm92.0$& $15.4\pm19.1$ & $0.102\pm0.071$\\  
25 & $21.8\pm23.9$ & $405.2\pm87.9$& $11.0\pm8.3$ & $0.081\pm0.065$\\  \midrule  
23 & $14.5\pm9.9$ & $542.7\pm153.4$& $12.9\pm6.6$ & $0.130\pm0.090$\\  
24 & $11.7\pm7.8$ & $443.9\pm94.2$& $17.4\pm21.4$ & $0.102\pm0.072$\\  
25 & $22.3\pm25.7$ & $409.4\pm89.5$& $11.3\pm8.8$ & $0.082\pm0.063$\\  \midrule  
23 & $13.2\pm9.3$ & $509.0\pm123.8$& $11.9\pm5.4$ & $0.125\pm0.087$\\  
24 & $11.8\pm10.2$ & $439.7\pm87.2$& $9.3\pm5.4$ & $0.101\pm0.070$\\  
25 & $18.7\pm8.4$ & $382.3\pm76.3$& $8.9\pm3.7$ & $0.076\pm0.076$\\  
\bottomrule
\end{tabular}
\caption{Sheath average values according to the different solar cycles and heliocentric distance. All distances (first block), inner heliosphere, r$<$ 1AU (second block) and outer heliosphere, r$\ge$ 1AU (third block).}
\label{tab:sheath_sc}
\end{table}

\begin{table}[H]
\begin{tabular}{ccccc}
\toprule
SC &N (cm$^{-3}$)& v (km/s) & B (nT) & Size (AU)\\  \toprule
23 & $7.3\pm4.9$ & $499.7\pm125.0$& $11.2\pm5.4$ & $0.336\pm0.331$\\  
24 & $6.5\pm4.2$ & $426.3\pm81.1$& $16.1\pm18.3$ & $0.252\pm0.156$\\  
25 & $13.4\pm12.5$ & $396.2\pm80.0$& $13.2\pm10.0$ & $0.177\pm0.114$\\  \midrule  
23 & $7.4\pm5.0$ & $505.9\pm127.2$& $12.9\pm6.6$ & $0.309\pm0.171$\\  
24 & $6.6\pm4.3$ & $426.1\pm82.3$& $17.4\pm21.4$ & $0.244\pm0.152$\\  
25 & $14.1\pm13.3$ & $397.6\pm80.4$& $11.3\pm8.8$ & $0.175\pm0.112$\\  \midrule  
23 & $7.1\pm4.8$ & $489.3\pm120.9$& $10.6\pm4.6$ & $0.306\pm0.175$\\  
24 & $6.4\pm4.2$ & $426.5\pm78.5$& $9.0\pm4.3$ & $0.269\pm0.161$\\  
25 & $9.4\pm5.2$ & $388.5\pm79.3$& $9.7\pm3.7$ & $0.186\pm0.127$\\  
\bottomrule
\end{tabular}
\caption{MO Cat I average values according to the different solar cycles and heliocentric distance. All distances (first block), inner heliosphere, r$<$ 1AU (second block) and outer heliosphere, r$\ge$ 1AU (third block).}
\label{tab:moI_sc}
\end{table}

\begin{table}[H]
\begin{tabular}{ccccc}
\toprule
SC &N (cm$^{-3}$)& v (km/s) & B (nT) & Size (AU)\\  \toprule
23 & $6.7\pm5.0$ & $450.3\pm87.5$& $7.8\pm4.9$ & $0.308\pm0.172$\\  
24 & $6.0\pm3.5$ & $398.6\pm65.6$& $9.9\pm8.4$ & $0.212\pm0.133$\\  
25 & $10.0\pm6.7$ & $376.0\pm52.1$& $10.9\pm6.5$ & $0.177\pm0.099$\\  \midrule  
23 & $7.9\pm4.8$ & $446.4\pm86.0$& $9.6\pm4.6$ & $0.276\pm0.144$\\  
24 & $6.1\pm3.4$ & $397.1\pm59.3$& $10.8\pm9.7$ & $0.210\pm0.137$\\  
25 & $9.8\pm7.0$ & $379.5\pm53.3$& $11.1\pm6.9$ & $0.186\pm0.102$\\  \midrule  
23 & $5.3\pm4.9$ & $455.2\pm89.4$& $5.7\pm4.5$ & $0.407\pm0.407$\\  
24 & $5.8\pm3.6$ & $401.2\pm75.6$& $7.7\pm3.2$ & $0.216\pm0.128$\\  
25 & $10.7\pm4.9$ & $353.9\pm39.6$& $9.4\pm3.1$ & $0.124\pm0.057$\\  
\bottomrule
\end{tabular}
\caption{MO Cat II average values according to the different solar cycles and heliocentric distance. All distances (first block), inner heliosphere, r$<$ 1AU (second block) and outer heliosphere, r$\ge$ 1AU (third block).}
\label{tab:moII_sc}
\end{table}

The sheath results from previous authors about the evolution with solar cycles, agree on the decreasing speed and magnetic field from SC23 to SC24. The speed of the sheaths according to our results for SC23 to SC24 (see Table \ref{tab:sheath_sc}) agree also with a decreasing speed. On the other hand, \citet{Yermolaev_2021} show an increase in density from SC23 to SC24 while \citet{Gopalswamy_2015_2} obtained a decrease in density. Our results show a decrease in density for all distance ranges studied, considering either the entire heliocentric distances, or separately only the inner heliosphere or only the outer heliosphere. \citet{Yermolaev_2021} and \citet{Gopalswamy_2015_2} show a decrease in magnetic field from SC23 to SC24, nevertheless, we obtained an increase in magnetic field, except for the outer heliosphere, however the standard deviation of some results for SC24 are high. The sheath size decreases from SC23 to SC24, although these values are slightly lower than those from \citet{Gopalswamy_2015_2}. The trend in SC25 reinforce the idea of a decreasing speed, magnetic field and size. Nevertheless, the density follows the increasing trend from the previous solar cycles. These trends for SC25 should be treated carefully since we have a small dataset for this cycle.

\citet{Yermolaev_2021} analyze separately the physical characteristics of sheaths and ejecta. They obtained almost the same density for SC23 and SC24. The density of MO Cat II agrees with this result, but only considering the outer heliosphere dataset. Considering either the inner heliosphere or the entire dataset, we obtain higher values with a decreasing trend. Something similar occurs with the magnetic field, we observe an increasing trend while they obtain similar values even with a small decrease. The speed trend is decreasing, agreeing with the results from \citet{Yermolaev_2021}. The size of MO Cat II with the solar cycles, shows a decreasing trend that seems to continue in SC25. 

The MO Cat I (see Table \ref{tab:moI_sc}) show almost the same trend as the MO Cat II (see Table \ref{tab:moII_sc}). A decrease in density from SC23 to SC24 followed by an increase in SC25, although the values are slightly higher. The same occurs with the speed, a decreasing trend with higher values than MO Cat II. The size of both MO Cat I and Cat II decrease around $50$\% from SC23 to SC25. Lastly, values for the magnetic field strength, although the values in SC24 have a high standard deviation, suggest a decrease in the inner and outer heliosphere, while considering the entire dataset, a small increase appears. 

\subsection{Magnetic obstacle Cat I and sheath ratio}
 \label{sec:mo_sheath_ratio}
\begin{figure}[H]
	\includegraphics[width=\columnwidth]{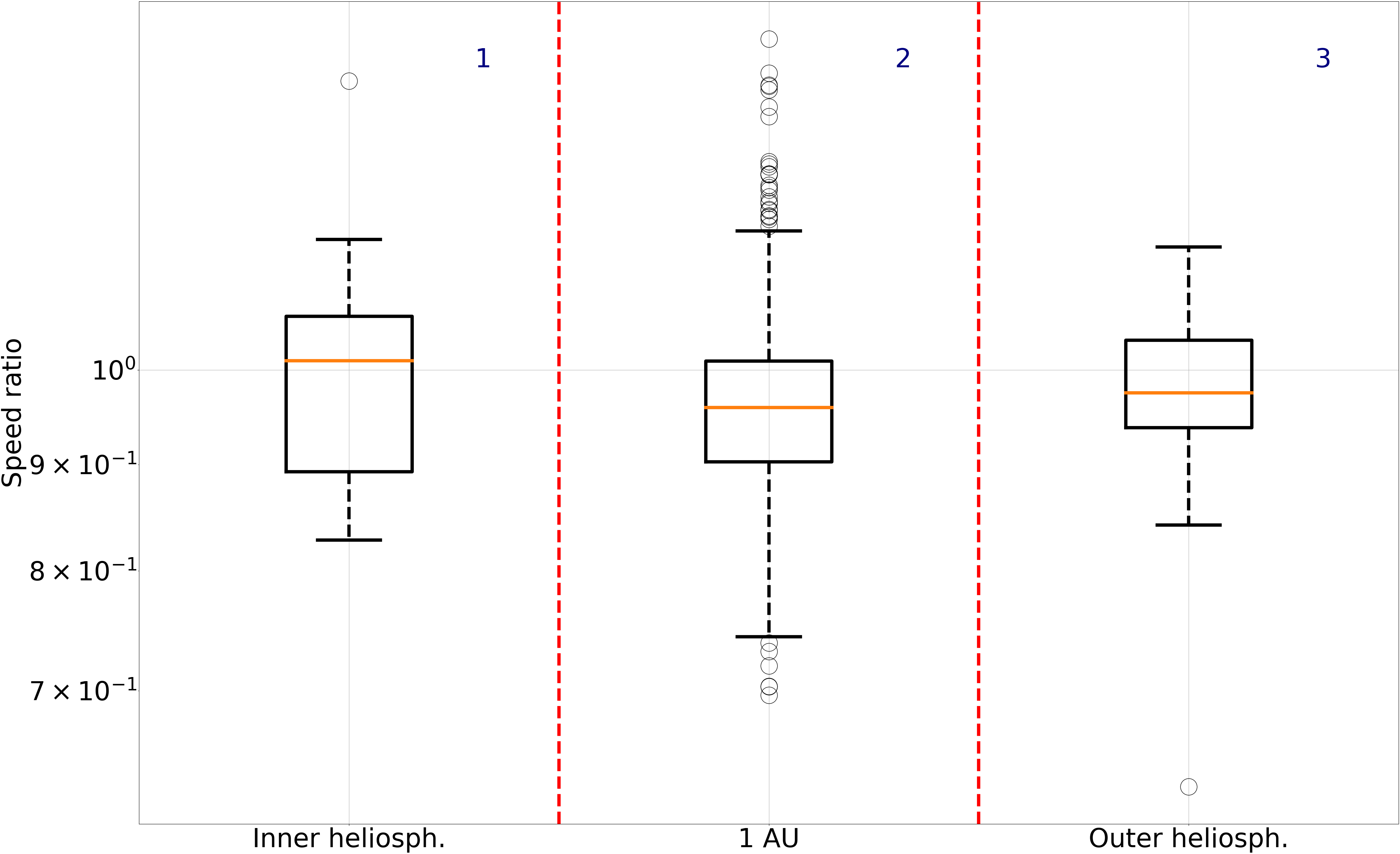}  
  \caption{Speed ratio between MO Cat I and sheath for different distance interval.}
  \label{fig:mo_sh_ratio_speed}
\end{figure}

\begin{figure}[H]
	\includegraphics[width=\columnwidth]{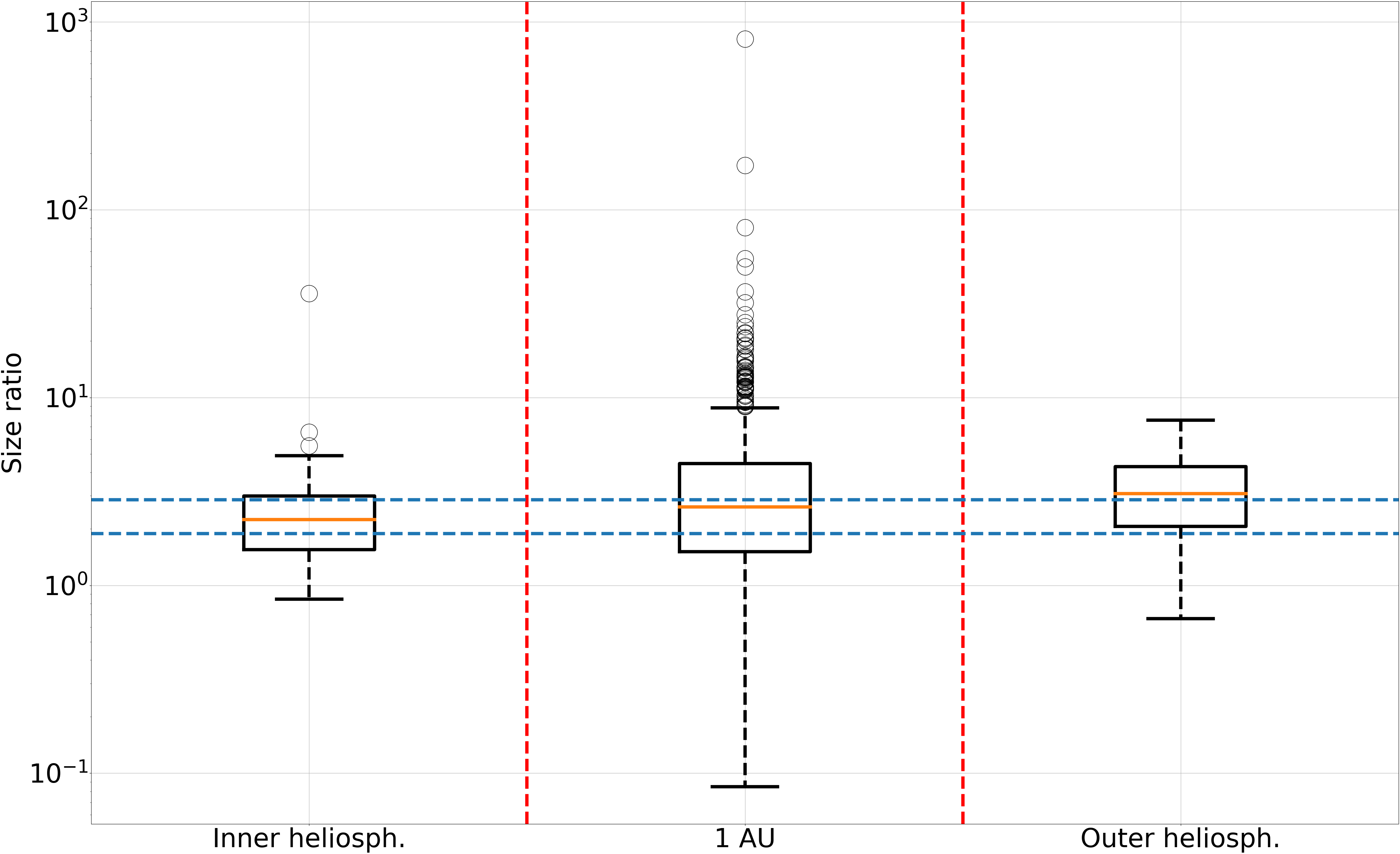}  
  \caption{Size ratio between MO Cat I and sheath for different distance interval. The blue dashed lines are the size ratio for the propagation and expansion type according to \citet{Salman_2021}.}
  \label{fig:mo_sh_ratio_size}
\end{figure}

In this section, we study the ratio between the sheaths and MO Cat I, i.e. the magnetic obstacle that drives and follows the sheath. For this, we split the heliocentric distance in three parts, r$<0.95$ AU, r$\in[0.95,1.05]$ AU and r$>1.05$ AU and defined the ratio as $\rm RT=x_{ MOI}/x_{SH}$, with $x$ as size or speed. Figure \ref{fig:mo_sh_ratio_speed} and \ref{fig:mo_sh_ratio_size} show the speed and size ratio respectively. The numbered block represents the result for each  distance interval: inner heliosphere, 1 AU and outer heliosphere.

Figure \ref{fig:mo_sh_ratio_speed} suggests that the speed ratio decreases from the inner to the outer heliosphere. In the inner heliosphere, it is slightly higher than 1, drops to 0.95 at 1 AU and increases to 0.97 in the outer heliosphere. In the three cases, the IQR is close to 1 suggesting that both structures have a similar speed.

With respect to the size ratio of the structures, MO Cat I are bigger than the sheaths over the entire distance interval. Indeed, Figure \ref{fig:mo_sh_ratio_size} shows that the size ratio increases by $\sim$15\% in each distance interval. \citet{Salman_2021} propose by analyzing 106 sheath regions near 1 AU using data from STEREO, a categorization of the sheaths according to their formation process: i) pure propagation-dominated sheaths, where solar wind tends to flow around the MO at a distance that is related to the relative speed between the MO and the solar wind; ii) pure expansion-dominated sheaths, where solar wind gets continuously piled-up over the entire extent of the structure due to the increasing size of the expanding MO. They obtain a size ratio between the magnetic obstacle and the sheath of 1.9 and 2.9 for the propagation and expansion type of formation respectively, represented by the blue-dashed lines in Figure \ref{fig:mo_sh_ratio_size}. Although the pure propagation or pure expansion formation mechanisms are uncommon, our results suggest that in the inner heliosphere there is a combination of both mechanisms, while moving to 1 AU and in the outer heliosphere, the dominant mechanism seems to be the expansion.

We further derive and discuss the correlation between the physical magnitudes of the sheath and MO Cat I. \citet{Salman_2020a} analyze the correlation coefficient (CC) of around 150 ICMEs seen by STEREO. In their work, they divide the ICMEs in three different categories according to the presence of sheath and shock. They obtain a CC of 0.91 between the speed of the magnetic ejecta and the sheath, and a CC of 0.64 between the magnetic fields of both structures. It should be noted that these values are for the ICMEs with sheath and shock, although their results for ICMEs only with sheath are very similar. Table \ref{tab:correl coefficient 4} shows the CC and its statistical significance according to the p-value test. The CC in parentheses are those found to be lower than 0.5 and statistically not significant, whereas the CC in brackets, are those lower than 0.5 but statistically significant. We find very strong correlations for the speed of both structures (CC \#1) over all the distance interval, with CC$\sim$0.9 in the three regions, which agrees with the results from \citet{Salman_2020a}. The correlation between the magnetic fields (CC \#2) is also high, especially in the inner and outer heliosphere. At 1 AU, the correlation is found to be moderate, weaker than the derived by \citet{Salman_2020a}. For the other parameters we get, at least for one of the distance intervals, a CC lower than 0.5. It should be noted that the number of ICMEs is not homogeneously distributed over the different distance intervals. The highest number is at 1 AU (983) followed by the inner heliosphere (177) and the outer heliosphere (83). Therefore, the CCs obtained for the outer heliosphere should be treated carefully since the number of events is low, even if they are statistically significant, e.g., CC \#8 and CC \#9. There is a strong correlation found between MO Cat I speed and the sheath temperature of the sheath (CC \#3) in the inner and at 1 AU. CC \#4, between the MO Cat I magnetic field and the sheath temperature, shows strong and statistically relevant CCs only in the inner heliosphere. CC \#5 and \#6 reveal very strong CCs only in the outer heliosphere. Nevertheless, in the inner heliosphere and at 1 AU, these CCs are moderate or weak, suggesting that they would not be relevant. On the other hand, the CC \#7, \#8 and \#9 show moderate CCs especially in the inner heliosphere and at 1 AU, suggesting a relevant correlation. Lastly, CC \#10 is the only example with a moderate anti-correlation in the inner heliosphere, while at 1 AU and in the outer heliosphere, there is no relevant correlation found.

\begin{table}[H]
\begin{tabular}{ccccc}
\toprule
\# & Parameter &r$<$0.95 AU & r $\in$[0.95,1.05] AU & r$>$1.05 AU\\
\toprule
1&v$_{MO I}$,v$_{Sh}$&0.86&0.91&0.88\\
2&B$_{MO I}$,B$_{Sh}$&0.80&0.51&0.82\\
3&v$_{MO I}$,T$_{Sh}$&0.70&0.80&(0.10)\\
4&B$_{MO I}$,T$_{Sh}$&0.78&[0.15]&0.65\\
5&T$_{MO I}$,T$_{Sh}$&(0.44)&[0.45]&0.93\\
6&T$_{MO I}$,B$_{Sh}$&(0.38)&[0.25]&0.87\\
7&v$_{MO I}$,B$_{Sh}$&0.50&0.56&(0.13)\\
8&N$_{MO I}$,N$_{Sh}$&0.66&[0.46]&[0.43]\\
9&B$_{MO I}$,N$_{Sh}$&0.60&0.53&[0.41]\\
10&B$_{MO I}$,Dur$_{Sh}$&-0.54&[-0.13]&(-0.22)\\
\bottomrule
\end{tabular}
\caption{Correlation coefficients (CC) between the physical magnitudes of sheath and MO Cat I. The values in parentheses show CCs below 0.5 and statistically not significant, while the values in brackets are below 0.5 and statistically significant.}
\label{tab:correl coefficient 4}
\end{table}

\section{Summary and Discussion} \label{sec:conclusions}

In this research we have gathered and compiled the results from 13 individual catalogs to create a dataset of more than 2000 separate ICMEs. These ICMEs are classified in ICME I and ICME II according to the presence or not of a sheath, allowing to define three structures, sheaths, MO Cat I and MO Cat II (see Figure \ref{fig:structures}). Due to the big sample obtained, the distance range covers the inner heliosphere (0.25 AU) to the outer heliosphere (5.42 AU) over the time range 1975--2022. 

\subsection{General evolution}
The size evolution of the magnetic obstacles is found to be similar for both categories, but MO Cat I is on average derived to be larger than MO Cat II over the distance range 0.75--1.25~AU. Investigating the evolution of the MO Cat II with distance, we derive only for the inner heliosphere a result consistent with a self-similar expansion (S$\propto$R) as defined by \citet{Farrugia_1993, Demoulin_2009a}. The MO Cat II evolution from the inner to the outer heliosphere may be affected by some mechanism like reconnection and "pancaking effect" \citep{Ruffenach_2015} which produce a non-self-similar expansion. 
For both MO categories, we derive the largest increase in size around 0.75 AU. Indeed, \citet{Scolini_2021} through the combined analysis of CME modelling and in-situ data of a halo CME in July 2012, found a phase of rapid expansion up to $\sim$0.4 AU in response to the rapidly decreasing pressure in the ambient solar wind \citep{Demoulin_2009a}. The sheath size evolution follows a similar trend, revealing a clear increase around 0.75 AU. Sheath size results at 1 AU from this study are found to be three times bigger than reported in previous studies \citep{Temmer_2022}, which might reflect differences in the sample used. While the sizes between MO Cat I and MO Cat II are different at 1AU, their densities are very similar and show a similar evolution with heliocentric distance. As expected, the sheaths reveal on average a higher density than the magnetic obstacles, being at 1 AU twice as dense as MO Cat I. In general, the density evolution with heliocentric distance of the three structures is very similar. The steepest density decrease is located around 0.75 AU, clearly related to the strong size enhancement at that distance.

\subsection{Sheath formation and relation to the magnetic obstacle}
The identification of ICMEs with clear sheath structures allows to statistically investigate the interaction between the magnetic obstacle that drives and follows the sheath over several distance ranges. 
Some works have shown the relation between the magnetic obstacle and the sheath, e.g. \citet{Masias-Meza_2016, Janvier_2019}. 
The speed ratio between MO Cat I and sheath in the inner heliosphere is slightly higher than 1 showing that the magnetic obstacle actively drives the sheath. 
At 1 AU and over the outer heliosphere, this ratio drops slightly below 1 suggesting that, at this distance the sheath is no longer driven by the magnetic obstacle and might be more comparable to a freely propagating distortion. This would coincide with the shifting of the solar wind density peak further upstream as derived for the outer heliosphere. 
 \citet{Salman_2021} proposed two sheath formation mechanisms: propagation-dominated, where the solar wind largely flows around the obstacle and expansion-dominated where the solar wind gets continuously piled-up all around the object \citep{Siscoe_2008}. According to the results from these authors, in the inner heliosphere there seems to be a balance between the contribution of both formation mechanisms. Nevertheless, moving outwards, the median size ratio between MO Cat I and sheath increases, suggesting that the contribution of the expansion mechanism would be more relevant. In our study, we find that the size of the MO Cat I increases by about 15\% in comparison with the size of the sheaths. The sheath formation mechanism therefore might change with the heliocentric distance, according to the interplay between MO increasing in size and decreasing in speed. However, further investigation analyzing in detail the expansion profile would be needed to confirm the different sheath formation mechanisms at work at different distances from the Sun. 

The correlation coefficient (CC) between several parameters of sheath and MO Cat I structures has also been studied. Comparing our results with those from \citet{Salman_2020a}, performed at 1 AU, we confirm a high correlation between the speed of the magnetic obstacle and the sheath (CC=0.91). In their work, they also found a strong correlation between the magnetic field of both structures (CC=0.64), but our results provide a lower CC of 0.51. The dataset used in this study allows extending the CC calculation to the inner and outer heliosphere. We note that the CC from the outer heliosphere, although being statistically significant according to the p-value test, should be treated carefully since the sample is small in comparison with the other distance ranges. The CC between speeds in the inner and outer heliosphere, is also high, with values of 0.86 and 0.88 respectively, showing that the correlation is high along the entire heliocentric distances. On the other hand, the CC of the magnetic fields in the inner (0.80) and outer heliosphere (0.82), increase in comparison to 1 AU. Other physical magnitudes have a relevant value (CC>0.5), see Table \ref{tab:correl coefficient 4}. In the inner heliosphere and at 1 AU, we find relevant CC between the speed of the magnetic obstacle and the temperature of the sheath and with the magnetic field of the sheath, and also between the magnetic field of the MO with the density of the sheath.

\subsection{Solar wind Interaction}
ICMEs are embedded in the solar wind and interact with the background solar wind during propagation. To investigate changes in the interaction processes such as compression, we have analyzed the relation between the upstream solar wind and the structures that immediately follow, i.e., sheaths and MO Cat II. Density ratios are calculated over a 4$\times$12-hours time window prior to the start time of the respective structure. The sheath density ratio suggests that in the inner heliosphere between 0.25 and 0.7 AU, the upstream solar wind is denser near the interface between solar wind and sheath. With farther heliospheric distance, the density accumulation gradually decreases, and between 1 and 1.5 AU, the densest region shifts upstream moving away from the interface between solar wind and sheath. In contrast to the results for the sheaths, in the inner heliosphere the MO Cat II does not show a clear trend below 0.7 AU. Between 0.7 and 1 AU, there is a small decrease, suggesting an increase in density in the nearest region to the interface between solar wind and MO Cat II. Above 1.5 AU, our results suggest that the central region of the upstream solar wind under study would have higher density. These results should be treated carefully, since the IQR of these boxplots are compatible with 1, meaning that the upstream solar wind and the MO Cat II could have the same density, i.e., are almost balanced.

\subsection{Solar cycle relations}
Our dataset covers almost five solar cycles, but for statistical reasons we focus the analysis on solar cycles SC23, SC24 and SC25 (rising phase). \citet{Kilpua_2017} provide the evolution with SC23 and part of the SC24 for sheaths and ICME at 1 AU. We extend these results to the inner and outer heliosphere, and in the inner heliosphere, the number of sheaths (the same as for MO Cat I) doubles from SC23 to SC24, and the same occurs with MO Cat II. The difference between MO Cat II with MO Cat I or sheaths remains constant for SC23 and SC24, with around 50\%. The magnetic field of MO Cat II shows lower values than MO Cat I and sheaths, this is most significantly revealed around the rising and maximum phase of SC24. In general, the solar wind speed of the structures (sheath, MO Cat I and MO Cat II) exhibits an increase after the maximum of each solar cycle, with the most significant one during SC23. While sheaths and MO Cat I show on average larger values compared to the yearly averaged solar wind speed, MO Cat II structures have much lower speeds, falling even below the ambient solar wind. Unlike the previous magnitudes, the evolution of density and size in the inner heliosphere does not show a clear relationship with the solar cycles. As expected, the sheaths are denser than the magnetic obstacles in both solar cycles, while their sizes are smaller. In the outer heliosphere, the number of sheaths (the same as for MO Cat I) in comparison with MO Cat II is very similar in some phases. Nevertheless, in SC24 the number doubles around the maximum phase, but it is half of what is observed in the inner heliosphere. The magnetic field strength of sheaths and MO Cat I suggests a moderate anti-correlation with the solar cycle, since there is an increase during the declining phase. On the other hand, the trend for the MO Cat II is less clear, with similar values to the other structures during some years. 
The comparison with previous results (e.g. \citet{Mitsakou_2014,Gopalswamy_2015_2}) of the evolution in SC23 and SC24 of sheaths, have shown an agreement with the trend in speed and size. Previous results for the sheath magnetic field have shown a decrease with solar cycles \citep{Gopalswamy_2015_2, Yermolaev_2021}, although our results show an increase but with a high standard deviation. There is no clear trend for the sheath density, although we have obtained a decrease with solar cycles for SC23 and SC24, whereas for SC25 there is an increase but with high standard deviation. The density of MO Cat I is higher than MO Cat II and these values agree with the results from \citet{Gopalswamy_2015_2}, specially considering our results in the inner heliosphere. The trend obtained for the magnetic field shows an increase for the solar cycles studied, whereas the previous results show an increase. On the other hand, our results show a decrease in speed, which was also obtained from previous studies \citep{Gopalswamy_2015_2,Yermolaev_2021}.

\section{Conclusions}\label{Conclusion}
The results obtained in this work provide important statistics on the evolution of different ICME structures (sheath, MO Cat I and MO Cat II) and the upstream solar wind for the inner heliosphere, but specially for the outer heliosphere, a distance range not fully covered yet. Our results show an abrupt change around 0.75 AU for size and density in the three different structures studied. Further analysis from spacecraft data for that specific distance range would be of high interest.

Although the sheath formation mechanism is still under debate, we have compared our results with those proposed by \citet{Salman_2021}, who classify the sheaths as either propagation-dominated or expansion-dominated sheaths. Applying that concept we have shown that the sheath formation mechanisms might change with the heliocentric distance, from a more propagation-dominated into a more expansion-dominated one. We also highlight the relation between the sheath and the magnetic obstacle, stating the importance of sheaths in the formation and evolution of magnetic obstacles. Since sheaths are the direct interface to the ambient solar wind, the relation between the upstream solar wind and the structures of the ICMEs are found to be important and should be studied further to enhance our understanding of ICME propagation and evolution in general. 

Results from this study provide valuable insights for CME model development, which could be useful to produce more accurate forecasts, helping to improve space weather tools and alerts.

\section{Acknowledgements}
C. Larrodera acknowledges the program 'Ayudas para la recualificación del sistema universitario español - UAH. Modalidad Margarita Salas 2021-2023' through the Spanish Ministry of Universities. 
Disclosure of Potential Conflicts of Interest: The authors declare that there are no conflicts of interest.

\bibliographystyle{apalike}  
\bibliography{references}  
\begin{appendix} 
\section{Catalogs description}  \label{app:cat_references}

Table \ref{tab:catalog_references} details the references of the individual catalogs used to obtain the combined catalog analyzed in this research. The column named 'Sheath' specifies if the catalog considers the presence of sheaths or not.

\begin{table}[H]
\begin{tabular}{ccc}
\toprule
\textbf{Catalog} & \textbf{Sheath}&  \textbf{Reference} \\ \toprule
1 &Yes &\citet{Mostl_2020}\\
 && \href{https://helioforecast.space/icmecat}{DOI: 10.6084/m9.figshare.6356420} \\ 
2 &Yes& \citet{Catalog_2}\\ 
3 &Yes& \href{https://parker.gsfc.nasa.gov/ICME_catalogs/STEREO/ICME_catalog_viewer_STEREO.php}{STEREO ICME List}\\ 
4 & Yes&\href{https://parker.gsfc.nasa.gov/ICME_catalogs/MESSENGER/ICME_catalog_viewer_MSG.php}{MESSENGER ICME List}\\ 
5 & Yes&\href{https://parker.gsfc.nasa.gov/ICME_catalogs/Helios/ICME_catalog_viewer_Helios.php}{Helios ICME List}\\ 
6 & Yes &\href{https://parker.gsfc.nasa.gov/ICME_catalogs/SO/ICME_catalog_viewer_SO.php}{Solar Orbiter ICME List}\\ 
7 & Yes&\href{https://parker.gsfc.nasa.gov/ICME_catalogs/PSP/ICME_catalog_viewer_PSP.php}{Parker Solar Probe ICME List}\\ 
8 &Yes& \citet{Helcats} \\ 
 && DOI: 10.6084/m9.figshare.4588315.v1 \\ 
9 &Yes& \citet{Regnault_2020} \\ 
10 &Yes& \citet{Catalog_10a} \\ 
 && \citet{Catalog_10b} \\ 
11 &Yes& \citet{Jian_2018} \\ 
12 &Yes& \citet{Jian_2006} \\ 
13 &No& \citet{Catalog_13} \\
\bottomrule
\end{tabular}
\caption{References of the individual catalog used.}
\label{tab:catalog_references}
\end{table}

As the different catalogs partly cover the same ICMEs, we removed all duplicates. For that, we were cross-checking catalogs from single spacecraft measurements with those from multiple spacecraft (catalog \#1 and \#8). 
\begin{enumerate}
    \item Set the time range associated with each ICME entry in the catalogs from a single spacecraft.
    \item Determine whether the time range identified in step 1 intersects with any ICME time ranges in the base catalog.
    \item If there is no overlap as described in step 2, we include this ICME entry in our final catalog.
\end{enumerate}

After combining the catalogs, the repeated events have been removed, and we obtain the 2136 separate ICMEs, from which 2003 belongs to the in-ecliptic ICMEs, i.e., latitude between $-$10º and $+$10º. The individual contributions to the final ICME catalog are detailed in Table \ref{tab:catalog_contrib}. The second and third column are the \% contributions to the entire final ICME catalog and the in-ecliptic ICME catalog, respectively. As can be noted, the main difference is the decrease of the ULYSSES ICMEs catalog when considering the in-ecliptic ICMEs.

\begin{table}[H]
\begin{tabular}{ccc}
\toprule
\textbf{Catalog} & \textbf{Entire catalog \%} & \textbf{In-ecliptic catalog \%}\\ \toprule
1 & 53.93& 57.31\\
2 & 1.17& 1.25\\
3 & 0.47& 0.50\\
4 & 0& 0\\
5 & 4.45& 4.74\\
6 & 0.23& 0.25\\
7 & 0.05& 0.05\\
8 & 1.31& 1.40\\
9 & 22.52& 24.01\\
10 & 2.34& 2.50\\
11 & 0.61& 0.65\\
12 & 4.49& 4.79\\
13 & 8.43& 2.55\\
\bottomrule
\end{tabular}
\caption{\% of contribution from the individual catalogs to the entire catalog (second column) and to the entire catalog considering only the in-ecliptic ICMEs.}
\label{tab:catalog_contrib}
\end{table}

The contribution from each spacecraft is detailed in Table \ref{tab:sc_contrib}, separated for the entire catalog and the in-ecliptic ICME catalog.

\begin{table}[H]
\begin{tabular}{ccc}
\toprule
\textbf{Catalog} & \textbf{Entire catalog \%} & \textbf{In-ecliptic catalog \%}\\ \toprule
Wind&27.95&29.81\\
ACE&24.86&26.51\\
STEREO-A&12.97&13.83\\
ULYSSES&8.66&2.60\\
STEREO-B&7.21&7.69\\
Helios&4.45&4.74\\
VEX&4.35&4.64\\
MESSENGER&4.07&4.34\\
PSP&1.69&1.80\\
Solar Orbiter&1.82&1.95\\
BepiColombo&1.50&1.60\\
MAVEN&0.47&0.50\\
\bottomrule
\end{tabular}
\caption{Percentage (\%) of contribution from each spacecraft to the entire catalog (second column) and to the entire catalog considering only the in-ecliptic ICMEs.}
\label{tab:sc_contrib}
\end{table}

The ICME signatures used for detection in the individual catalogs are very similar for all the catalogs. The only catalogs that does not consider the sheaths are catalog \#10 and \#13, as explained below. 

The complete catalog obtained from this research can be found in this \href{https://edatos.consorciomadrono.es/dataset.xhtml?persistentId=doi:10.21950/XGUIYX}{link}.

\section{Literature review from previous authors} 
\subsection{Size evolution with heliocentric distance} \label{app:size}

The power law fitting parameters of the size obtained from previous authors are detailed in Table \ref{tab:references_size}. \citet{Chen_1996} (R1) obtained the fitting parameters from theoretical calculations between 0.3 and 5 AU, while \citet{Kumar_1996} (R2) provide the size of ICMEs between and 0.3 and 4 AU by comparing with observations.
\citet{Bothmer_1998} (R3) using magnetic clouds detected by Helios 1 and 2 estimate their size within the range between 0.3 and 4.2 AU. 
\citet{Wang_2005} (R4) and \citet{Liu_2005} (R5) complemented the analysis of the magnetic clouds with ICMEs measured by Ulysses and ACE in both cases. \citet{Liu_2005} (R5) also add ICMEs measured from Wind, while \citet{Wang_2005} (R4) add others from Pioneer Venus Orbiter (PVO). In both cases, the distance range was extended to 5.4 AU.
\citet{Leitner_2007} (R6.1) extended even more the distance range to 10 AU, thanks to the measurements from Pioneer 10 and 11, and Voyager 1 and 2. They also estimate the size of the inner heliosphere using measurements from Helios and Wind (R6.2)
\citet{Gulisano_2010} estimate the size of magnetic clouds in the inner heliosphere (between 0.3 and 1 AU) measured by Helios (R7.1), separating them between those who have not been affected by other ICMEs or fast stream from a coronal hole, known as non-perturbed MC (R7.2), or those affected, known as perturbed magnetic obstacles (R7.3).
\citet{Gulisano_2012} following the research from the previous work, analyze the size in the outer heliosphere (between 1.4 and 5.4 AU) using measurements from Ulysses, for the entire dataset (R8.1), non-perturbed (R8.2) and perturbed (R8.3).
Finally, \citet{Temmer_2022} obtained the size of magnetic obstacle (R9.1) and sheaths (R9.2) in the inner heliosphere (between 0.3 and 1 AU) using measurements from Helios and Parker Solar Probe. 

\begin{table}[H]
\begin{tabular}{ccc}
\toprule
Reference & Dist. range (AU) & Size \\ 
\toprule
R1 & 0.3-5 & $S= r^{0.88}$ \\
R2 & 0.3-4 & $S= 0.148 \cdot r^{0.97}$ \\
R3 & 0.3-4.2 & $S=\left ( 0.24\pm0.01\right) \cdot r^{\left(0.78\pm0.1\right)}$ \\
R4 & 0.3-5.4 & $S=0.19 \cdot r^{0.61}$ \\
R5 & 0.3-5.4 & $S=\left ( 0.25\pm0.01\right) \cdot r^{\left(0.92\pm0.07\right)}$ \\
R6.1 & 0.3-10 & $S=\left ( 0.195\pm0.017\right) \cdot r^{\left(0.61\pm0.087\right)}$ \\
R6.2 & 0.3-1 & $S=\left ( 0.23\pm0.05\right) \cdot r^{\left(1.14\pm0.44\right)}$ \\
R7.1 & 0.3-1 & $S=r^{\left(0.78\pm0.12\right)}$ \\
R7.2& 0.3-1 & $S=r^{\left(0.89\pm0.15\right)}$ \\
R7.3& 0.3-1 & $S=r^{\left(0.45\pm0.16\right)}$ \\

R8.1 & 1.4-5.4 & $S=r^{\left(0.56\pm0.34\right)}$ \\
R8.2 & 1.4-5.4& $S=r^{\left(0.79\pm0.46\right)}$ \\
R8.3 & 1.4-5.4 & $S=r^{\left(0.54\pm0.48\right)}$ \\
R9.1 & 0.3-1 & $S=0.27 \cdot r^{0.78}$ \\
R9.2 & 0.3-1 & $S=0.04 \cdot r^{0.48}$ \\
\bottomrule
\end{tabular}
\caption{Overview about the power law size fitting from previous authors. The references are detailed in the text.}
\label{tab:references_size}
\end{table}

\subsection{Density evolution with heliocentric distance} \label{app:density}

The summary of the density power law fitting parameters is detailed in Table \ref{tab:references_density}. \citet{Kilpua_2017} (R10) from measurements of Helios 1 and 2, Wind, ACE Ulysses and Voyager extend the power law fitting until 12 AU. As detailed in the previous section, R8.1, R8.2 and R8.3 refers to estimation of the density using all the magnetic clouds, only the non-perturbed and the perturbed respectively from \citet{Gulisano_2012}, while R9.1 and R9.2 refers to the density of magnetic ejecta and sheaths respectively from \citet{Temmer_2022}. 

\begin{table}[H]
\begin{tabular}{ccc}
\toprule
Reference & Dist. range (AU) & Density \\ 
\toprule
R2 & 0.3-4 & $N= 7.2 \cdot r^{-2.8}$ \\
R3 & 0.3-4.2 & $N=\left (6.47\pm0.85\right) \cdot r^{\left(-2.4\pm0.3\right)}$ \\
R4 & 0.3-5.4 & $N=6.7 \cdot r^{-2.4}$ \\
R5 & 0.3-5.4 & $N=\left ( 6.16\pm0.3\right) \cdot r^{\left(-2.32\pm0.07\right)}$ \\
R6.1 & 0.3-10 & $N=\left ( 6.63\pm0.28\right) \cdot r^{\left(-2.62\pm0.07\right)}$ \\
R6.2 & 0.3-1 & $N=\left ( 7.24\pm1.51\right) \cdot r^{\left(-2.44\pm0.46\right)}$ \\

R8.1 & 1.4-5.4 & $N=r^{\left(-1.7\pm0.43\right)}$ \\
R8.2 & 1.4-5.4& $N=r^{\left(-2.24\pm0.66\right)}$ \\
R8.3 & 1.4-5.4 & $N=r^{\left(-1.4\pm0.5\right)}$ \\
R9.1 & 0.3-1 & $N=7.1 \cdot r^{-2.4}$ \\
R9.2 & 0.3-1 & $N=22.3 \cdot r^{-1.7}$ \\
R10 & 0.3-12 & $N=\left (5.74\pm0.27\right) \cdot r^{\left(-2.21\pm0.03\right)}$ \\
\bottomrule
\end{tabular}
\caption{Overview about the power law density fitting from previous authors. The references are detailed in the text.}
\label{tab:references_density}
\end{table}

\subsection{Evolution with solar cycle} \label{app:sc_evol}

About the evolution with the solar cycle, \citet{Mitsakou_2014} from the analysis of 325 ICMEs with sheaths obtained from a dataset of OMNI near-Earth database and defined by their in situ plasma signatures during SC23 (between 1996 and 2008), obtained the average values of different magnitudes, detailed in Table \ref{tab:mitsakou_sc_icmes}, where the first block are the results for the sheath and the second for the ejecta.

\begin{table}[H]
\begin{tabular}{ccccc}
\toprule
SC &N (cm$^{-3}$)& v (km/s) & B (nT) & Size (AU)\\  \toprule
23 & $12.1\pm0.5$ & $504\pm7$& $11.3\pm0.3$ & $0.111\pm0.005$\\ \midrule
23 & $6.8\pm0.2$ & $467\pm6$& $10.1\pm0.2$ & $0.27\pm0.01$\\ 
\bottomrule
\end{tabular}
\caption{Average values from \citet{Mitsakou_2014} of sheath (first block) and ejecta (second block).}
\label{tab:mitsakou_sc_icmes}
\end{table}

\citep{Gopalswamy_2015_2} analyzed sheaths and magnetic clouds from the SC23 and part of the SC24 (between May 1996 to December 2014) from in-situ measurements at the L1 point. The results are detailed in Table \ref{tab:gopalswamy_sc} where the first block is for the sheaths and the second block for the magnetic clouds.

\begin{table}[H]
\begin{tabular}{ccccc}
\toprule
SC &N (cm$^{-3}$)& v (km/s) & B (nT) & Size (AU)\\  \toprule
23 & $15.25$ & $539.5$& $19.18$ & 0.145\\  
24 & $12.90$ & $447.4$& $13.24$ & 0.139\\ \midrule  
23 & $7.549$ & $473.9$& $16.54$ & 0.224\\  
24 & $6.897$ & $402.1$& $12.33$ &0.174\\ 
\bottomrule
\end{tabular}
\caption{Average values from \citet{Gopalswamy_2015_2} of sheath (first block) and magnetic clouds (second block).}
\label{tab:gopalswamy_sc}
\end{table}

Another study, shows the results from \citet{Yermolaev_2021}, where through the analysis of OMNI data from SC21 to SC24, obtained the average value of the characteristic magnitudes of the solar wind, sheath and ejecta during these cycles (Table \ref{tab:yermolaev_sc_icmes}).

\begin{table}[H]
\begin{tabular}{cccc}
\toprule
SC &N (cm$^{-3}$)& v (km/s) & B (nT) \\  \toprule
23 & $11.2\pm8.3$ & $470\pm120$& $10.3\pm6.1$ \\  
24 & $12.4\pm9.4$ & $430\pm100$& $8.3\pm4.5$ \\  \midrule
23 & $5.2\pm4.5$ & $450\pm100$& $6.6\pm3.1$ \\  
24 & $5.4\pm3.8$ & $410\pm80$& $6.0\pm2.5$ \\  
\bottomrule
\end{tabular}
\caption{Average values from \citet{Yermolaev_2021} of sheath (first block) and ejecta (second block).}
\label{tab:yermolaev_sc_icmes}
\end{table}

\end{appendix}

\end{document}